\documentclass[12pt,preprint]{aastex}
\usepackage{natbib}
\newcommand\cm{{\,\rm cm}}
\newcommand\pcc{{\,\rm cm}^{-3}}
\newcommand\K{{\;\rm K}}
\newcommand\erg{{\;\rm erg}}
\newcommand\yr{{\;\rm yr}}
\newcommand\Myr{{\;\rm Myr}}
\newcommand\Msun{{\;\rm\,M_\odot}}
\newcommand\kms{{\;\rm km\; s^{-1}}}
\newcommand\pc{{\;\rm\,pc}}
\newcommand\kpc{{\;\rm kpc}}
\newcommand\simgt{\lower.5ex\hbox{$\; \buildrel > \over \sim \;$}}
\newcommand\simlt{\lower.5ex\hbox{$\; \buildrel < \over \sim \;$}}
\newcommand\epsff{{\; \varepsilon_{\rm ff}}}

\newcommand\siggbc{\Sigma_{\rm GBC}}
\newcommand\sigdiff{\Sigma_{\rm diff}}
\newcommand\sigsfr{\Sigma_{\rm SFR}}
\newcommand\Pth{P_{\rm th}}
\newcommand\vth{v_{\rm th}}
\newcommand\fwt{\tilde f_w}
\newcommand\tsfgbc{t_{\rm SF, GBC}}
\newcommand\tsf{t_{\rm SF}}
\newcommand\fdiff{f_{\rm diff}}
\newcommand\zdiffmax{z_{\rm diff,max}}
\newcommand\rhodm{\rho_{\rm dm}}
\newcommand\rhoext{\rho_{\rm sd}}

\slugcomment{Accepted for publication in the Ap.J.}

\shorttitle{Regulation of Star Formation}
\shortauthors{Ostriker et al}

\begin{document}


\title{Regulation of Star Formation Rates in Multiphase Galactic Disks: a Thermal/Dynamical Equilibrium Model}


\author{Eve C. Ostriker\altaffilmark{1}, Christopher F. McKee \altaffilmark{2},
and Adam K. Leroy\altaffilmark{3}}
\altaffiltext{1}{Department of Astronomy, University of Maryland, College Park, MD 20742} 
\altaffiltext{2}{Departments of Physics and Astronomy, University of California, Berkeley, CA 94720; 
and 
%Laboratoire d'Etudes du Rayonnement et de la Mati\'ere en Astrophysique, 
LERMA-LRA, Ecole Normale Superieure, 24 rue Lhomond, 75005
                            Paris, France}
\altaffiltext{3}{Hubble Fellow; National Radio Astronomy Observatory, 520 Edgemont Road, Charlottesville, VA 22903}
\email{ostriker@astro.umd.edu,cmckee@astro.berkeley.edu,aleroy@nrao.edu}

\begin{abstract}
We develop a model for the regulation of galactic star formation rates
$\sigsfr$ in disk galaxies, in which ISM heating by stellar UV plays a
key role.  By requiring that thermal and (vertical)
dynamical equilibrium are simultaneously satisfied within the diffuse
gas, and that stars form at a rate proportional to the mass of the
self-gravitating component, we obtain 
a prediction for $\sigsfr$ as a function of 
the total gaseous surface density $\Sigma$ and the
midplane density of stars + dark matter $\rhoext$.  
The physical
basis of this relationship is that the thermal pressure in the diffuse
ISM, which is proportional to the UV heating rate and therefore to
$\sigsfr$, must adjust until it matches the midplane pressure value
set by the vertical gravitational field. 
Our model applies to regions where $\Sigma \simlt 100 \Msun \pc^{-2}$.  
In low-$\sigsfr$
(outer-galaxy) regions where diffuse gas dominates, the theory
predicts that $\sigsfr \propto \Sigma \sqrt{\rhoext}$.   
The decrease of thermal equilibrium pressure when $\sigsfr$ is low implies, 
consistent with
observations, that star formation can extend (with declining efficiency) 
to large radii in
galaxies, rather than having a sharp cutoff at a fixed value of
$\Sigma$.  The main parameters entering our model are the ratio of
thermal pressure to total pressure in the diffuse ISM, the fraction of
diffuse gas that is in the warm phase, and the star formation timescale in
self-gravitating clouds; all of these are (at least in principle)
direct observables.  At low surface density, our model depends on 
the ratio of the mean midplane FUV intensity (or thermal pressure in 
the diffuse gas) to the star formation rate, which we set based on Solar
neighborhood values.
We compare our results to recent observations,
showing good agreement overall for azimuthally-averaged data in a set
of spiral galaxies.  For the large flocculent spiral galaxies NGC 7331 
and NGC 5055,
the correspondence between theory and observation is remarkably close.
\end{abstract}


\section{Introduction}

Star formation is regulated by many physical factors, with processes
from sub-pc to super-kpc scales contributing to setting the overall
rate 
(see e.g. \citealt{MO07}).  
One of the key factors expected to control the star formation rate is the
available supply of gas.  Over the
whole range of star-forming systems, from entire spiral galaxies to
circumnuclear starbursts, the global average of the surface density of
star formation, $\sigsfr$, is observed to be correlated with the
global average of the neutral gas surface density $\Sigma$ as
$\sigsfr\propto\Sigma^{1+p}$  with $1+p\approx 1.4$ \citep{Ken98}.
Recent observations at high spatial resolution have made it possible
to investigate local, rather than global, correlations of the star
formation rate with $\Sigma$, using either azimuthal
averages over rings, or mapping with apertures down to $\simlt \kpc$
scales
(e.g. \citealt{2002ApJ...569..157W,2003MNRAS.346.1215B,2007ApJS..173..524B,
  2004ApJ...602..723H,
  2005PASJ...57..733K,2007A&A...461..143S,2007ApJ...671..333K,
  2008AJ....136..479D,Bigiel08,2009ApJ...704..842B,2010A&A...510A..64V}).  
While power-law \citep{1959ApJ...129..243S,1963ApJ...137..758S}
relationships are still evident in these local
studies, steeper slopes are found for the outer, atomic-dominated
regions of spiral galaxies (as well as  dwarf galaxies) 
compared to the inner, molecular-dominated
regions of spirals.  
In addition, measured indices in the low-$\Sigma$, low-$\sigsfr$ regime 
vary considerably from one galaxy to another.  Thus, no single Schmidt law
characterizes the regulation of star formation on local scales in 
the outer parts of galaxies.

The nonlinearity of observed Schmidt laws implies that not just the
quantity of gas, but also its physical state and the surrounding
galactic environment, affect the star formation rate.  Indices $p>0$
imply that the star formation efficiency is higher in higher-density
regions, which are generally nearer the centers of galaxies and have
shorter dynamical times.  Indeed, the expectation based on theory and
numerical simulations \citep{GoldreichLB65,KO01} in thin, single-phase
gaseous disks is that gravitational instabilities leading to star
formation would grow only if the Toomre parameter \citep{Toomre64}
$Q\equiv \vth \kappa/(\pi G \Sigma)$ is sufficiently small.  These
instabilities would develop over a timescale comparable to the
galactic orbital time $t_{\rm orb}\equiv 2\pi /\Omega$, which
corresponds to about twice the two-dimensional Jeans time
$t_{J,2D}\equiv \vth/(G\Sigma)$ when $Q$ is near-critical.  Here,
$\vth$ is the thermal speed ($\vth^2\equiv \Pth/\rho=k T/\mu$ for
$\Pth$, $\rho$, and $T$ the gas thermal 
pressure, density, and temperature), and $\kappa$ is the epicyclic
frequency ($\kappa^2\equiv R^{-3} d\Omega^2/dR$).  While the implied
scaling $\sigsfr \propto \Sigma \Omega$ is roughly satisfied globally
\citep{Ken98}, supporting the notion that galaxies evolve towards
states with $Q$ roughly near-critical
(e.g. \citealt{1972ApJ...176L...9Q}), for more local observations this
does not provide an accurate prediction of star formation
(e.g. \citealt{Ler08,2009ApJ...705..650W}).

In addition to galactic rotation and shear rates, an important aspect of local
galactic environment 
is the gravity of the 
stellar component.  The background stellar gravity compresses the disk 
vertically (affecting the three-dimensional Jeans time 
$t_J\equiv \sqrt{\pi/(G \rho)}$), and perturbations in the stellar density
can act in concert with gaseous perturbations in gravitational instabilities 
(altering the effective $Q$).  Thus, 
one might expect the stellar surface density $\Sigma_s$ and/or volume 
density $\rho_s$ to affect the star formation rate
(see e.g. \citealt{2007ApJ...660.1232K} and references therein).  
For example, if the 
stellar vertical gravity dominates that of the gas (see \S 2 for a detailed 
discussion of this), a scaling $\sigsfr\propto \Sigma/t_J$ would imply 
$\sigsfr \propto \Sigma^{3/2} (G^3 \rho_s)^{1/4}/\vth$ for a constant-temperature
gas disk.  Although star formation does appear to be correlated with $\Sigma_s$ 
(e.g. \citealt{1994ApJ...430..142R,1998ApJ...493..595H}; see also below), 
the simple scaling  $\propto \Sigma/t_J$ 
(taking into account both gaseous and 
stellar gravity, and assuming constant $\vth$, in calculating $t_J$) 
does not in fact provide an accurate
local prediction of star formation rates 
(e.g. \citealt{2008ARep...52..257A,Ler08,2009ApJ...705..650W}).  

A likely reason for the inaccuracy of the simple star formation
prescriptions described above is that they do not account for the
multiphase character of the interstellar medium (ISM), in which most
of the volume is filled with low-density warm (or hot) 
gas but much (or even most) of
the mass is found in clouds at densities two or more orders of
magnitude greater than that of the intercloud medium.  For the colder
(atomic and molecular) phases, the turbulent velocity dispersions are
much larger than $\vth$, so that the mean gas density $\bar \rho$
averaged over the disk thickness depends on the turbulent vertical
velocity dispersion.  Even when multiphase gas and turbulence
(and stellar and gas gravity) are taken into account in simulations,
however, the simple
estimate $\sigsfr \propto \Sigma/t_J\propto \Sigma \sqrt{G \bar \rho}$ 
(using $\bar\rho$ directly measured from the simulations) yields
Schmidt-law indices steeper than the true values measured in both the
simulations and in real galaxies \citep{KO09a}.
Perhaps this should not be
surprising, since one would expect the proportions of gas among different
phases, as well as the overall vertical distribution, to affect the
star formation rate.  If, for example, most of the ISM's mass were in
clouds of fixed internal density that formed stars at a fixed rate,
then increasing the vertical velocity dispersion of this system of
clouds would lower $\bar\rho$ but leave $\sigsfr$ unchanged.

The relative proportions of gas among different phases seems difficult to
calculate from first principles, because it depends on how
self-gravitating molecular clouds form and how they are destroyed,
both of which are very complex processes.
Intriguingly, however, analysis of recent observations 
of spiral galaxies has shown that the
surface density of the molecular component averaged over $\sim \kpc$ annuli
or local patches shows a relatively simple overall behavior, 
increasing roughly linearly with the empirically-estimated 
midplane gas pressure \citep{2002ApJ...569..157W,BR04, BR06, Ler08}.  
The physical reason behind 
this empirical relation has not, however, yet been explained.  

In this paper, we use a simple physical model to analyze how the
gas is partitioned into diffuse and self-gravitating 
components, based on
considerations of dynamic and thermodynamic equilibrium.  
We develop the idea that the midplane pressure in the diffuse component must
simultaneously satisfy constraints imposed by vertical force balance, 
and by balance between heating (primarily from UV) 
and cooling. In particular, we propose that the
approximately linear empirical relation between molecular content and 
midplane pressure identified by \citet{BR04, BR06} arises because
the equilibrium gas pressure is approximately proportional to the 
UV heating rate; since the mean
UV intensity is proportional to the star formation rate and the star
formation rate is proportional to the molecular mass 
in normal spirals, the observed relationship naturally 
emerges.\footnote{\citet{1985ApJ...295L...5D} previously showed that 
assuming the pressure is proportional to the star formation rate yields 
scaling properties similar to observed relationships.}
We use our analysis to predict the dependence
of the star formation rate on the local gas, stellar, and 
dark matter content of disks, and
compare our predictions with observations.  The analysis, including our
basic assumptions and observational motivation for parameters that enter 
the theory, is set out in \S 2.  Section 3 then compares to the observed 
data set previously presented in \citet{Ler08}.  In \S 4, we summarize 
and discuss our main results.

\section{Analysis}

\subsection{Model concepts and construction
\label{section-outline}
}

In this section, we construct a local steady-state model for the star
formation rate in the disk, with independent variables the total
surface density of neutral gas ($\Sigma$), the midplane stellar
density ($\rho_s$), and the dark matter density ($\rho_{\rm dm}$).
The latter two quantities enter only through their effect on the
vertical gravitational field.  To develop this model, we suppose that
the diffuse gas filling most of the volume of the interstellar medium
is in an equilibrium state.  The equilibrium in the diffuse ISM has
two aspects: force balance in the vertical direction (with a sum of
pressure forces offsetting a sum of gravitational forces), and balance
between heating and cooling (where heating is dominated by the
FUV). Star-forming clouds, because they are self-gravitating entities
at much higher pressure than their surroundings, are treated as
separate from the space-filling diffuse ISM.  The abundance of
gravitationally-bound, 
star-forming clouds is nevertheless important for establishing an
equilibrium state in the diffuse gas, because the FUV that heats the
diffuse ISM originates in young OB associations.  We assume
(consistent with observations and numerical simulations) that the
equilibrium thermal state established for the diffuse medium includes
both warm and cold atomic gas.  This hypothesis leads to a connection
between the dynamical equilibrium state and the thermal equilibrium
state: there are two separate constraints on the pressure that must be
simultaneously satisfied.  
These conditions are met by an
appropriate partition of the available neutral gas
into diffuse and self-gravitating components.

The reason for the partition between diffuse and self-gravitating gas
can be understood by considering the physical requirements for
equilibrium.  The specific heating rate ($\Gamma$) in the diffuse gas
is proportional to the star formation rate, which is proportional to
the amount of gas that has settled out of the vertically-dispersed
diffuse gas and collected into self-gravitating clouds.  The specific
cooling rate ($n\Lambda$) in the diffuse gas is proportional to the
density and hence to the thermal pressure, which (if force balance holds) 
is proportional to
the vertical gravity and to the total surface density of diffuse gas.
Thus, an equilibrium state, in which cooling balances heating and
pressure balances gravity, can be obtained by a suitable division of
the gas mass into star-forming (gravitationally-bound) 
and diffuse components such that their
ratio is proportional to the vertical gravitational field.  If too
large a fraction of the total surface density is in diffuse gas, the
pressure will be too high, while the star formation rate will be too
low.  In this situation, the cooling would exceed heating, and mass
would ``drop out'' of the diffuse component to produce additional
star-forming gas. With additional star formation, the FUV intensity
would raise the heating rate in the diffuse gas until it matches the
cooling.

In the remainder of this section, we formalize these ideas
mathematically, first defining terms (\S \ref{section-components}),
then considering the requirements of dynamical balance (\S
\ref{section-momentum}) and thermal balance (\S \ref{section-energy}),
and finally combining these to obtain an expression for the star
formation rate when both equilibria are satisfied (\S
\ref{section-SF-solution}).  We then discuss, from a physical point of
view, how the various feedback processes might act to adjust the
system over time, steering it toward the equilibrium we have
identified (\S \ref{section-approach}).  While evolving to an
equilibrium of this kind is plausible, we emphasize that this is an
assumption of the present model, which must be tested by detailed
time-dependent simulations.\footnote{ Very recent numerical studies
  provide support for the quasi-equilibrium assumption -- see Kim,
  Kim, \& Ostriker (2010, in preparation).}  A worked example applying
the model to an idealized galaxy is presented in \S
\ref{section-ideal}.  In developing the present model, we have adopted
a number of simplifications that a more refined treatment should
address; we enumerate several of these issues in \S
\ref{sec-consider}.

\subsection{Gas components
\label{section-components}
}

In this model, we divide the neutral ISM into two components. One component
consists of the gas that is collected into gravitationally-bound
clouds (GBCs) localized near the galactic midplane, with mean surface density
(averaged over $\sim$ kpc scales) of $\siggbc$.  The other component
consists of gas that is diffuse (i.e. not gravitationally bound), with
mean surface density $\sigdiff$.  Here, we use the term ``diffuse'' in 
the sense of being widely dispersed or scattered throughout the volume; 
the diffuse component may include both tenuous, volume-filling 
gas and small, dense cloudlets (see below). 
All star formation is assumed to take place within the GBC component.
In normal galaxies, the GBC component is identified with the population of 
giant molecular clouds (GMCs). 
Note that while observed GMCs in the Milky Way consist primarily of 
molecular gas, they also contain atomic gas in shielding layers. More 
generally, as we shall discuss
further below, the relative proportions of molecular and dense atomic gas in 
GBCs depends on the cloud column and metallicity, and GBCs could even be 
primarily atomic if the metallicity is sufficiently low.

The diffuse component is identified (in normal galaxies) with the atomic ISM. 
We treat the diffuse gas 
as a two-phase cloud-intercloud medium in thermal pressure equilibrium,
with turbulent vertical velocity dispersion $v_z^2$ assumed to be the
same for warm and cold phases. 
Although the cold cloudlets within the diffuse component have much 
higher internal
density than the warm intercloud gas, they are (by definition)
each of sufficiently 
low mass that they are non-self-gravitating, such that their thermal pressure
(approximately) matches that of their surroundings. 
The pressure in the interior of 
GBCs is considerably higher than the pressure of 
the surrounding diffuse gas (cf. \citealt{KO09b}). 

In reality, the diffuse gas would not have a single unique pressure
even if the radiative heating rate is constant
because of time-dependent dynamical   
effects: turbulent compressions and rarefactions
heat and cool the gas, altering what would otherwise be a balance
between radiative heating and cooling processes.  Nevertheless, simulations 
of turbulent gas with atomic-ISM heating and cooling  
indicate that the majority of the gas has pressure within $\sim 50\%$ of 
the mean value \citep{PO05,PO07}, although the breadth of the 
pressure peak depends on the timescale of turbulent forcing 
$\sim L_{\rm turb}/v_{\rm turb}$ compared to the cooling time 
\citep{2005A&A...433....1A,2010A&A...511A..76A,2007A&A...465..431H,2005ApJ...630..911G,2009ApJ...693..656G,2006ApJ...653.1266J,2009ApJ...704..137J}.
Observations indicate a range of pressures in the cold atomic gas in 
the Solar neighborhood, with a small fraction of the gas 
at very high pressures, and $\sim 50\%$ of the gas at pressures 
within $\sim 50\%$ of the mean value \citep{Jen01,2007IAUS..237...53J}.

In general, 
the volume-weighted mean thermal pressure at the midplane is given by 
\begin{eqnarray}
\langle \Pth \rangle_{\rm vol} 
&=&\frac{\int \Pth d^3 x  }{\int d^3 x  }
=\frac{\int(\Pth/\rho) \rho d^3x}{\int d^3x}
=\frac{\int \rho d^3x}{\int d^3x} \frac{  \int \vth^2 dm}{\int dm  }  
\nonumber\\
&=&\rho_0 \langle \vth^2 \rangle_{\rm mass},
\label{Pth_vol}
\end{eqnarray}
where $\rho_0$ is the volume-weighted mean midplane density
of diffuse gas.
The quantity $\langle \vth^2 \rangle_{\rm mass}$ is the mass-weighted 
mean thermal velocity dispersion; for a medium with warm and 
cold gas with respective mass fractions (in the diffuse component) 
$f_w$ and $f_c=1-f_w$ and temperatures $T_w$ and $T_c$,
\begin{equation}
\frac{\langle \vth^2 \rangle_{\rm mass}  }{c_w^2  }
= f_w + \frac{ T_c}{T_w}(1-f_w) \equiv \fwt.
\label{vth-eq}
\end{equation}
Here, $c_w\equiv (P_w/\rho_w)^{1/2}=(k T_w/\mu)^{1/2}$ is the 
thermal speed of warm gas.
Since the ratio $T_w/T_c$ is typically 
$\sim 100$, $\fwt \approx f_w$ unless 
$f_w$ is extremely small.

If the thermal pressure in the warm and cold diffuse gas phases are the same, 
$\langle \Pth \rangle_{\rm vol} = P_w=\rho_w c_w^2$, so that from equations 
(\ref{Pth_vol}) and (\ref{vth-eq}),
\begin{equation}
\frac{\rho_w}{\rho_0} = 
\frac{\langle \vth^2 \rangle_{\rm mass}  }{c_w^2  }=\fwt.
\label{massfrac-eq}
\end{equation}
This result still holds approximately even if the warm and cold medium
pressures differ somewhat, since the warm gas fills most of the 
volume, $\langle \Pth \rangle_{\rm vol} \approx P_w$.
Note that one can also write 
$\rho_w/\rho_0=f_w (V_{\rm tot}/V_w)$ for $V_{\rm tot}$ and $V_w$ the total 
and warm-medium volumes, so that $\fwt \approx f_w$ provided the warm medium
fills most of the volume. If the medium is all cold gas, 
$\fwt = T_c/T_w$.  Henceforth, we shall assume the warm and cold gas
pressures are equal at the midplane so that 
$\langle \Pth \rangle_{\rm vol} \rightarrow \Pth$; for convenience, we 
shall also omit the subscript on $\langle \vth^2 \rangle_{\rm mass}$.

\subsection{Vertical dynamical equilibrium of diffuse gas
\label{section-momentum}
}

By averaging the momentum equation of the diffuse component
horizontally and in time, and integrating outward from the midplane,
it is straightforward to show that the difference in the total vertical momentum
flux across the disk thickness 
(i.e. between midplane and $\zdiffmax$)
must be equal to the total weight of the diffuse gas 
(e.g. \citealt{1990ApJ...365..544B,PO07,KO09b}).  This total
weight has three terms.  The first term is 
the weight of the diffuse gas in its own gravitational field, 
\begin{equation}
\int_0^{\zdiffmax} \rho \frac{d \Phi_{\rm diff}}{dz} dz
=\frac{1}{8\pi G} \int_0^{\zdiffmax} 
\frac{d\left(\frac{d \Phi_{\rm diff}  }{d z  }\right)^2 }
{d z  } dz 
=\frac {\pi G \sigdiff^2  } { 2 },
\label{gas_weight}
\end{equation}
where we have used 
$\vert d \Phi_{\rm diff}/dz\vert_{\zdiffmax} = 2 \pi G \sigdiff$
for a slab.  The second term is the 
weight of the diffuse gas in the mean gravitational field
associated with the GBCs, 
\begin{equation}
\int_0^{\zdiffmax} \rho \frac{d \Phi_{\rm GBC}}{dz} dz \approx
\pi G \siggbc \sigdiff,
\label{gas_weight_gbc}
\end{equation}
where we have assumed that the scale height of the 
GBC distribution
is much smaller than that of the diffuse gas so that 
$\vert d \Phi_{\rm GBC}/dz\vert \approx 2 \pi G \siggbc$ over most of the 
integral. Note that equation (\ref{gas_weight_gbc}) gives 
an upper bound on this term in the weight, with a lower bound 
given by $\pi G \siggbc \sigdiff/2$, corresponding to the case in 
which the vertical 
distributions of the diffuse and gravitationally-bound components are the
same.  
The third term is the weight in
the gravitational field associated with the disk stars plus dark
matter, 
\begin{equation}
\int_0^{\zdiffmax} \rho \left(\frac{d \Phi_{\rm s}}{dz} + 
\frac{d\Phi_{\rm dm}}{dz} \right)dz \equiv
2 \pi \zeta_d G \frac{\rhoext \sigdiff^2}{\rho_0}.
\label{gas_weight_*}
\end{equation}
Here, $\rhoext=\rho_s+\rho_{\rm dm}$
is the midplane density of the stellar disk 
plus that of the dark matter halo; 
we have assumed a flat rotation curve $V_c=const.$ for 
the dark halo so that 
$\rhodm= (V_c/R)^2/(4\pi G)$.\footnote{If the vertical stellar 
distribution in the disk 
varies as $\rho_s \propto {\rm sech}^2(z/H_s)$ with 
$H_s=v_{z,s}^2/(\pi G \Sigma_s)$, then the midplane stellar density is 
$\rho_s=\Sigma_s/(2H_s)=\pi G \Sigma_s^2/(2 v_{z,s}^2)$.  
Existing photometric and kinematic 
observations suggest that $H_s\sim const$ and $v_{z,s}\propto \Sigma_s^{1/2}$ 
\citep{1982A&A...110...61V,1993A&A...275...16B}, 
but these are sensitive primarily to the 
central parts of the disk.  Note that 
if the Toomre parameter for the stellar disk and the vertical-to-horizontal 
velocity dispersion ratio are both constant with radius, then 
$\rho_s \propto \rhodm\propto (V_c/R)^2$.  In this case, the ratio of the 
gas-to-stellar scale height is $\sim v_{z,g} /v_{z,s}$; since
the gas can dissipate turbulence and cool to maintain constant $v_{z,g}$ 
while the stellar velocity dispersion secularly increases 
over time, the gas layer will tend to be 
thinner than the stellar layer even if both components flare in the outer 
parts of galaxies.} 
The stellar disk's scale height is assumed to be larger than that of
the diffuse gas, so that $g_z\approx 4\pi G \rhoext z$ within the
diffuse gas layer.  The numerical value of $\zeta_d$ depends, but not
sensitively, on the exact vertical distribution of the gas, which in
turn depends on whether self- or external gravity dominates; $\zeta_d
\approx 0.33$ within $5\%$ for a range of cases between zero external
gravity and zero self gravity.  Allowing for a gradient in the vertical
stellar density within the gas distribution, the stellar contribution 
to the weight would be reduced by a factor $\sim 1- (2/3) (H_g/H_s)^2$, where 
$H_g$ and $H_s$ are the gaseous and stellar scale heights.
In the (unlikely) circumstance that 
the diffuse-gas scale height is much larger than that of the stars, 
$g_z\approx 2 \pi G \Sigma_s$  would be substituted for the gravity 
of the stellar component, yielding a contribution analogous to that in 
equation (\ref{gas_weight_gbc}) with $\siggbc \rightarrow \Sigma_s$.

Including both thermal and kinetic terms, and taking 
$\rho \rightarrow 0$ at the top of the diffuse-gas layer, the difference in
the gaseous vertical
momentum flux between $z=0$ and $\zdiffmax$
is given by $\Pth + \rho_0 v_z^2$.  
The term $v_z^2$ is formally a mass-weighted quantity (analogous to 
$\langle \vth^2\rangle_{\rm mass}$), but we assume a similar 
turbulent velocity dispersion for the diffuse warm and cold atomic gas 
\citep{HT03}.
If the magnetic
field is significant, a term equal to the difference between 
$B^2/(8\pi) - B_z^2/(4\pi)$ at $z=0$ and $\zdiffmax$ is added 
\citep{1990ApJ...365..544B,PO07}.  Like other pressures,
these magnetic terms are volume-weighted; both observations 
\citep{HT05} and numerical simulations \citep{PO05} indicate 
that field strengths in the warm and cold atomic medium are similar.
If the scale height of the magnetic field is larger than that
of the diffuse gas (as some observations indicate; see e.g. \citealt{Fer01}), 
then this term will be small, while it will provide an appreciable effect 
if $B\rightarrow 0$ where $\rho\rightarrow 0$.  
In any case, the magnetic term in
the vertical momentum flux may be accounted for by taking 
$\rho_0 v_z^2\rightarrow \rho_0 v_z^2 + \Delta B^2/2 - \Delta B_z^2\equiv 
\rho_0 v_t^2$, 
where 
% $v_A\equiv B/\sqrt{4\pi\rho_0}$ is the total (effective) 
%Alfv\'en speed and 
%$v_{A,z}\equiv B_z/\sqrt{4\pi\rho_0}$, and 
$\Delta$ indicates the difference
between values of the squared magnetic field 
at $z=0$ and $\zdiffmax$.  Cosmic rays have a much larger
scale height than that of the diffuse (neutral) gas, such that difference
in the cosmic ray pressure between $z=0$ and $\zdiffmax$ may be
neglected.  We have also neglected the contribution from diffuse warm
ionized gas, which has a low mean density and large scale height compared 
to that of the neutral gas (e.g. \citealt{2008PASA...25..184G}).

Equating the momentum flux difference with the total weight, we have
\begin{equation}
\Pth\left(1+ \frac{v_t^2}{c_w^2 \fwt}\right)
=\frac{\pi G }{2}\sigdiff^2 + 
\pi G \siggbc \sigdiff +
2  \pi \zeta_d G c_w^2 \fwt  \frac{\rhoext\sigdiff^2}{\Pth}. 
\label{mom-eq}
\end{equation}
Here, we have used equations 
(\ref{Pth_vol}) and (\ref{vth-eq}) to
substitute $\fwt c_w^2/\Pth$ for $\rho_0^{-1}$ on the right-hand side.  
As noted above, the second term on the right-hand side
could be reduced by up to a factor of two, if the scale height of 
the GBC 
distribution approaches that of the 
diffuse gas.
It is convenient to define
\begin{equation}
\alpha\equiv 1 + \frac{v_t^2}{c_w^2\fwt}= 
\frac{\langle\vth^2\rangle + v_t^2   }{\langle \vth^2 \rangle  }
=\frac{\Pth + \rho_0 v_z^2 +  \Delta(B^2/2 - B_z^2)/(4\pi) }{\Pth},
\label{alpha_def}
\end{equation}
which represents the midplane 
ratio of total effective pressure 
to thermal pressure.  If the magnetic contribution is small (which would 
be true if $\Delta B^2 \ll B^2$, even if magnetic and thermal pressures are
comparable at the midplane), $\alpha$ is 
the total observed velocity dispersion $\sigma_z^2$ 
divided by the mean thermal value. We shall treat
$v_t$, $c_w$, and $\fwt$ 
as parameters that do not vary strongly within a galaxy 
or from one galaxy to another (see below), 
and $\sigdiff$, $\siggbc$, and $\Pth$ as (interdependent) variables.  
At any location in a galaxy, we shall consider $\rhoext$ 
(and the total gas surface density $\Sigma=\sigdiff + \siggbc$)
as a given environmental conditions.

Equation (\ref{mom-eq}) is a quadratic in both $\sigdiff$ and $\Pth$.
Thus, if $\Pth$ and $\siggbc$ are known, we may solve to obtain the
surface density of diffuse gas:
\begin{equation}
\sigdiff= \frac{2\alpha \Pth  }{\pi G \siggbc + \left[(\pi G \siggbc)^2 + 
2 \pi G \alpha (\Pth + 4\zeta_d  c_w^2 \fwt \rhoext)  \right]^{1/2}  }. 
\label{sigdiff-eq}
\end{equation}
Scaling the variables to astronomical units, the result in equation
(\ref{sigdiff-eq}) can also be expressed as
\begin{equation}
\sigdiff =  
\frac{9.5 \Msun\pc^{-2} \alpha \left(\frac{\Pth/k  }{3000 \K \pcc }  \right)  }
{ 0.11  \left(\frac{\siggbc  }{1 \Msun \pc^{-2}  }  \right) 
+ \left[0.011 \left(\frac{\siggbc  }{1 \Msun \pc^{-2}  }  \right)^2
+ \alpha\left(\frac{\Pth/k  }{3000 \K \pcc }  \right)
+ 10 \alpha \fwt\left(\frac{\rhoext  }{0.1 \Msun \pc^{-3}  }   \right)
 \right]^{1/2}}.
\label{sigdiff-dimen}
\end{equation}

What are appropriate parameter values to use?  Since thermal balance
in the warm medium is controlled by line cooling \citep{Wol95,Wol03},
the warm medium temperature is relatively insensitive to local
conditions in a galaxy; we shall adopt $c_w=8\kms$.  Numerical
simulations in multiphase gas have shown that the magnetic field is
amplified by the magnetorotational instability to a level
$B^2/(8\pi)=(1-2)\Pth$, independent of the mass fractions of cold and
warm gas and the vertical gravitational field strength
\citep{PO05,PO07}, while $|B_z/B_\phi|\ll 1$.  This is consistent with
observed magnetic field strengths measured in the Milky Way and in
external galaxies \citep{HT05,Beck08}.  Large-scale turbulent velocity
dispersions observed in local HI gas (both warm and cold) are $\sim
7\kms$, comparable to $c_w$ \citep{HT03,Mohan04}.  Total vertical
velocity dispersions in HI gas in external galaxies are also observed
to be in the range $5-15\kms$, decreasing outward from the center
\citep{Tam09}.  

The most uncertain parameter is $\fwt\approx f_w$, the fraction of the
diffuse mass in the warm phase.  In the Solar neighborhood, this is
$\sim 0.6$ \citep{HT03}, and in external galaxies the presence of both
narrower and broader components of 21 cm emission suggests that both
warm and cold gas are present \citep{deBlok06}, with some indication
based on ``universality'' in line profile shapes that the warm-to-cold
mass ratio does not strongly vary with position \citep{Pet07}.  In the
outer Milky Way, the ratio of HI emission to absorption appears nearly
constant out to $\sim 25$ kpc, indicating that the warm-to-cold ratio
does not vary significantly \citep{Dic09}.  In dwarf galaxies as well,
observations indicate that both a cold and warm HI component is
present \citep{1996ApJ...462..203Y}.  While uncertain, it
is likely that $\fwt\sim 0.5-1$, at least in outer galaxies.  

Thus, allowing for the full range of observed variation,
$\alpha\sim 2-10$, $\alpha \fwt \sim 1-5$, and $\alpha/\fwt \sim 2-20$; 
we shall adopt $\alpha=5$ and $\fwt=0.5$ as typical for mid-to-outer-disk 
conditions.
For these fiducial parameters, and 
taking midplane thermal pressure $P_{\rm th,0}/k \sim 3000 \K \pcc$
(see \citealt{Jen01} and
\citealt{Wol03}), $\rhoext=0.05 \Msun \pc^{-3}$ \citep{Hol00}, 
and $\siggbc \simlt 2  \Msun \pc^{-2}$ 
\citep{1987ApJ...322..706D,2001ApJ...547..792D,1988ApJ...324..248B,2006ApJ...641..938L,2006PASJ...58..847N}
near the Sun,
the result from equation (\ref{sigdiff-dimen}) is
consistent with the observed total surface density estimate
$\sim 10 \Msun \pc^{-2}$ 
of atomic gas in the Solar neighborhood 
\citep{Dic90,Kal09}. 

Equation (\ref{mom-eq}) may also be solved to obtain the thermal pressure in 
terms of $\sigdiff$, $\siggbc$, $\rhoext$ and the diffuse-gas parameters 
$\alpha$ and $\fwt$:
\begin{equation}
\Pth=\frac{\pi G \sigdiff^2}{4\alpha}
\left\{
{1 + 2\frac{\siggbc  }{\sigdiff  } +
\left[\left(1+ 2\frac{\siggbc  }{\sigdiff  }  \right)^2  
+\frac{32\zeta_d c_w^2\fwt \alpha  }{\pi G  }\frac{\rhoext}{\sigdiff^2 }\right]^{1/2}}
\right\}.
\label{pth-eq}
\end{equation}
Over most of the disk in normal galaxies, 
the term in equation (\ref{pth-eq}) that is
proportional to $\rhoext$ (arising from the weight in the  
stellar-plus-dark matter gravitational field) dominates; this yields
\begin{equation}
\Pth\sim \sigdiff \left(2 G \rhoext  \right)^{1/2} 
\left(\frac{\pi \zeta_d \fwt }{\alpha }  \right)^{1/2} c_w.
\label{approx-P}
\end{equation}
For given $\rhoext$ and $\Sigma$, the thermal pressure therefore
increases approximately proportional to the fraction of gas in the
diffuse phase, $\fdiff \equiv \sigdiff/\Sigma$.  With 
$\left(\pi \zeta_d  \right)^{1/2}\approx 1$ and $\fwt \approx f_w$,
$\Pth \sim \Sigma_w \left(2 G \rhoext  \right)^{1/2} c_w^2/
(\langle \vth^2\rangle + v_t^2)^{1/2}$ in this limit; i.e. it is the 
surface density of the volume-filling warm medium that sets the thermal 
pressure. 
Multiplying equation (\ref{approx-P}) by $\alpha$ and using 
$\alpha \fwt= (\langle \vth^2\rangle + v_t^2)/c_w^2$ 
yields
\begin{equation}
P_{\rm tot}\sim \sigdiff \left(2 G \rhoext  \right)^{1/2} 
(\langle \vth^2\rangle + v_t^2)^{1/2}.
\end{equation}
This is the 
same as the formula for midplane pressure
adopted by \citet{BR04,BR06}, except that instead of $\sigdiff$ their
expression contains the total gas surface density $\Sigma$, 
instead of $(\langle \vth^2\rangle + v_t^2)^{1/2}$
they use the thermal+turbulent vertical velocity dispersion 
(these are equal if vertical 
magnetic support is negligible), and they omit the 
dark matter contribution to $\rhoext$.
Using $\rho_s=\pi G
\Sigma_s^2/(2 v_{z,s}^2)$ and taking $\siggbc$, $\rhodm \rightarrow
0$, equation (\ref{pth-eq}) yields
\begin{equation}
P_{\rm tot} \sim \frac{\pi G \Sigma^2}{2}\left(1+ 
\frac{[\langle \vth^2\rangle + v_t^2)^{1/2}  ]^{1/2} \Sigma_s }{v_{z,s} \Sigma}\right),
\end{equation}
recovering the result of \citet{1989ApJ...338..178E} (except that 
he includes the total $B^2$, rather than $\Delta B^2$, in $v_t^2$ and 
$P_{\rm tot})$.

\subsection{Thermal equilibrium of diffuse gas
\label{section-energy}
}

As expressed by equation (\ref{pth-eq}), the thermal pressure in the
diffuse gas must respond to the dynamical constraint imposed by vertical
momentum conservation in the disk.  In addition, the thermal pressure is
also regulated by the microphysics of heating and cooling.  Namely, if
the atomic gas is in the two-phase regime (as is expected in a star-forming 
region of a galaxy;
see \S \ref{section-SF-solution}), then the thermal pressure
must lie between the minimum value for which a cold phase is possible,
$P_{\rm min,cold}$, and the maximum value for which a warm phase is possible, 
$P_{\rm max, warm}$.  \citet{Wol03} found, based on detailed modeling of 
heating and cooling 
in the Solar neighborhood, that  
geometric mean of these two equilibrium extrema, 
$P_{\rm two-phase}\equiv (P_{\rm min,cold} P_{\rm max, warm})^{1/2}$, 
is comparable to the local empirically-estimated thermal pressure,
and that two phases 
are expected to be present in the Milky Way out to $\sim 18\kpc$.  Based on
turbulent numerical simulations with a bistable cooling curve, 
\citet{PO05,PO07} found that the mean 
midplane pressure evolves to a value near the geometric mean pressure 
$P_{\rm two-phase}$, 
for a wide range of vertical gravitational fields and warm-to-cold mass 
fractions.
Thus, we expect the midplane thermal pressure in the diffuse gas 
to be comparable to the two-phase value
defined by the thermal equilibrium curve, 
$\Pth\approx P_{\rm two-phase}$.
Since $P_{\rm max, warm}/P_{\rm min,cold}\sim 2-5$ \citep{Wol03}, even if 
$\Pth=P_{\rm two-phase}$ 
does not hold precisely, the midplane 
pressure $\Pth$ will be within a factor $\sim 2$ of 
$P_{\rm two-phase}$ 
provided the 
diffuse gas is in the two-phase regime.

For the fiducial Solar neighborhood model of \citet{Wol03}, the geometric 
mean thermal pressure is 
$P_{\rm two-phase}/k 
\sim P_{\rm th,0}/k \sim 3000 \K \pcc
$. 
For other environments, 
the values of $P_{\rm min, cold}$ and  $P_{\rm max,warm}$ depend on the
heating of the gas: enhanced heating pushes the transition pressures
upward \citep{Wol95}.  
Because the dominant heating is provided by the photoelectric
effect on small grains, 
$P_{\rm two-phase}$ 
increases approximately linearly with the FUV
intensity.  Assuming that 
$P_{\rm two-phase}$ 
scales with $P_{\rm min,cold}$, 
we adapt the expression given in \citet{Wol03} and normalize  
$P_{\rm two-phase}$
using
the Solar-neighborhood value: 
\begin{equation}
\frac{P_{\rm two-phase}}{k  }= 12000 \K \pcc \frac{G_0' Z_d'/Z_g'  }
{1 + 3.1 (G_0' Z_d'/\zeta_t')^{0.365}}.
\label{Wol-eq}
\end{equation}
Here, $G_0'$ is the mean FUV intensity relative to the Solar
neighborhood value 
$J_{\rm FUV, 0}=2.1\times 10^{-4} \erg \cm^{-2} {\rm s}^{-1}{\rm sr}^{-1}$, 
$\zeta_t'$ is equal to the total cosmic ray/EUV/X
ray ionization rate relative to the value $10^{-16}{\rm s}^{-1}$, and
$Z_d'$ and $Z_g'$ are respectively the dust and gas abundances
relative to Solar-neighborhood values.  

The FUV is produced by OB associations, 
so that the intensity will be proportional to the star formation rate
per unit surface area.  The intensity also depends on the radiative
transfer of the UV.  For example, a slab of gas with UV optical depth
$\tau_\perp = \kappa \Sigma$ and total surface density of FUV emission
$\Sigma_{\rm FUV}$ has 
$J_{\rm FUV}=\Sigma_{\rm  FUV}(1-E_2(\tau_\perp/2))/(4\pi \tau_\perp)$, 
where $E_2$ is the second exponential integral.  
This correction for
radiative transfer depends logarithmically on $1/\tau_\perp$ at low
optical depth (yielding a variation $\propto R$ if $\Sigma$ is an
exponential -- see \citet{1994ApJ...435L.121E}), varying by $\sim 50\%$ 
for $\tau_\perp \sim 0.1-1$.  For simplicity, we neglect
variations in $J_{\rm FUV}$ associated with the radiative transfer
here; we shall simply assume $J_{\rm FUV}\propto \Sigma_{\rm FUV}\propto 
\sigsfr$. 
\citet{Fuchs2009} (see also \citealt{2001AJ....121.1013B,2002A&A...390..917V}) 
find that the 
Solar-neighborhood value of the surface star formation rate
$\Sigma_{\rm SFR,0}$ is $2.5\times 10^{-9}\Msun\pc^{-2}\yr^{-1}$, 
which then yields
\begin{equation}
{G_0'}\equiv \frac{J_{\rm FUV}}{J_{\rm FUV,0} }
\approx \frac{\Sigma_{\rm SFR}}{\Sigma_{\rm SFR,0}}
=\frac{\Sigma_{\rm SFR}  }{2.5 \times 10^{-9}\Msun\pc^{-2}\yr^{-1}  }.
\label{G0-eq}
\end{equation}
Since equation (\ref{Wol-eq}) is normalized using the observed 
Solar-neighborhood pressure $P_{\rm th,0}$ 
and equation (\ref{G0-eq}) is normalized 
using the (inverse of the) 
observed Solar-neighborhood star formation rate $\Sigma_{\rm SFR,0}$, 
our model depends only on the ratio of these quantities,
$P_{\rm th,0}/\Sigma_{\rm SFR,0}$.  Equivalently, 
since \citet{Wol03} predict the value of 
$P_{\rm th,0}/J_{\rm FUV,0}$ from theory, our model depends on 
the measured ratio of local FUV intensity to local star formation rate,
$J_{\rm FUV,0}/\Sigma_{\rm SFR,0}$.
Although here we adopt an empirical value for $J_{\rm FUV}/\sigsfr$ based 
on the Solar neighborhood, in principle this ratio
may be calculated theoretically with a detailed radiative transfer 
and population synthesis model.  Simple estimates using standard relationships
between the FUV emission and $\sigsfr$ (e.g. \citealt{2007ApJS..173..267S})
yield values of $J_{\rm FUV}/\sigsfr$ similar to our adopted value from 
the Solar neighborhood.

Assuming the high-energy ionization rate is proportional to the local 
value of $\sigsfr$ and inversely proportional to $\Sigma$ \citep{Wol03}, 
\begin{equation}
\frac{G_0'}{\zeta_t'}=\frac{\Sigma}{\Sigma_0}, 
\label{zet-eq}
\end{equation}
where $\Sigma_0$ is the surface density of 
neutral gas at the Solar circle.  Strictly speaking, the above would only 
apply to cosmic rays, with $\zeta_t' \propto \sigdiff^{-1}$ instead of 
$\Sigma^{-1}$ for soft X-rays and and EUV.  In practice, however, this does
not affect the results for the star formation rate, since the dependence
on this term is weak to begin with, and only enters the prediction for 
$\sigsfr$ in outer disks (see eq. \ref{asymp-SFR}), 
where $\Sigma \rightarrow \sigdiff$.  

Taking $\Sigma_0\sim 10\Msun\pc^{-2}$ and 
setting $Z_d'/Z_g'=1$ yields, for 
$\Pth=P_{\rm two-phase}$,
\begin{equation}
\sigsfr\approx 6 \times 10^{-10} \Msun \pc^{-2}\yr^{-1}
\left(\frac{\Pth/k  }{3000 \K \pcc } \right)
\left[1+ 3 \left(\frac{Z_d'\Sigma}{10 \Msun\pc^{-2}} \right)^{0.4}  \right]; 
\label{SFR-P-eq}
\end{equation}
this can be combined with equation (\ref{pth-eq}) to yield a prediction 
for the star formation rate in terms of the gas and stellar contents of
the disk. 
More generally, equation (\ref{G0-eq}) may be inserted into equation 
(\ref{Wol-eq}), and the result substituted in equation (\ref{sigdiff-eq}) or 
(\ref{pth-eq}) to obtain, respectively, 
an expression for $\sigdiff$ in terms of 
$\sigsfr$, $\rhoext$, and $\siggbc$, or an expression for $\sigsfr$ in terms
of $\sigdiff$, $\rhoext$, and $\siggbc$.  

\subsection{The equilibrium star formation rate
\label{section-SF-solution}
}

As our goal is to obtain a prediction for $\sigsfr$ in terms of the
total gaseous surface density $\Sigma = \sigdiff + \siggbc$ 
and midplane 
stellar+dark matter density $\rhoext$,
we require an 
additional relationship among the variables.  Since star formation is assumed 
to take place only within 
GBCs,
if the timescale 
to convert this gas to stars is $\tsfgbc$, then 
\begin{equation}
\sigsfr= \frac{\siggbc}{\tsfgbc}=\frac{\Sigma-\sigdiff}{\tsfgbc}.
\label{sfr-eq}
\end{equation}
In normal galaxies, GBCs are identified with GMCs (the outer layers of 
which are in fact atomic -- see below).
Recently, \citet{Bigiel08} found that there is an approximately linear
relationship between the molecular mass measured in CO and the star
formation rate, over the mid-disk regions in a set of spiral
galaxies where $\Sigma\sim 10 - 100 \Msun \pc^{-2}$. 
The measured proportionality constant is
 $t_{\rm SF, CO}\equiv \Sigma_{\rm mol,CO}/\sigsfr\approx
2\times 10^9 \yr$. 
Some recent studies targeting this regime in spirals at $\sim \kpc$ scales 
find weak systematic variations in $t_{\rm SF,CO}$ with surface density 
\citep{2002ApJ...569..157W,2004ApJ...602..723H,2007ApJ...671..333K,
2010A&A...510A..64V}. These are mild (no more than a factor of $2-3$ over 
the range $\Sigma = 10 - 100 \Msun \pc^{-2}$) and almost all find a consistent 
normalization, with 2 Gyr being a typical timescale. 
 CO is present only in the portions of clouds 
where $A_V\simgt 1$ \citep{2010ApJ...716.1191W,2010arXiv1003.1340G}, so 
that it can become a poor tracer of molecular mass in low-metallicity systems.
From observations of dust emission in the SMC, however, 
the star formation timescale in ``dark'' molecular gas (where hydrogen is 
in H$_2$ but carbon is atomic rather than in CO) is 
found to be similar to that in CO-bright gas 
(A. Bolatto, personal communication). Based on these empirical results,
$\tsfgbc = t_{\rm SF, mol} \siggbc/\Sigma_{\rm
  mol}$ for  $t_{\rm SF, mol} \sim const.$  

Since GBCs contain both dense atomic
shielding exteriors and dense molecular shielded interiors, while
molecular gas may be both within GBCs and in unbound clouds, we can write 
\begin{eqnarray}
\frac{\siggbc}{\Sigma_{\rm mol}}&=& 
\frac{M_{\rm mol,GBC} + M_{\rm atom,GBC}}{ 
M_{\rm mol,GBC} + M_{\rm mol,diff}}\nonumber\\
&=&
\frac{ 1 + (M_{\rm atom}/M_{\rm mol})_{\rm GBC}}
{1 + \Sigma_{\rm mol, diff}/\Sigma_{\rm mol, GBC}}.  
\end{eqnarray}
The atomic-to-molecular ratio within
externally-irradiated (spherical) 
clouds depends primarily on the metallicity and
total cloud column density of hydrogen 
$N_{\rm H,cloud}\equiv M_{\rm cloud}/(\mu_H \pi R_{\rm cloud}^2)$ as 
$(M_{\rm atom}/M_{\rm mol})_{\rm cloud}\approx [Z'^{0.8}
  (N_{\rm H,cloud}/1.8\times 10^{21}\cm^{-2})-0.7]^{-1}$ 
\citep{KMT09,2010ApJ...709..308M}. Here, $N_{\rm H}=N_{\rm HI} + 2 N_{\rm H2}$. 
Assuming the column densities of GBCs are similar to the observed values
$N_{\rm H,cloud}\sim  10^{22}\cm^{-2}$ 
(i.e. $\Sigma_{\rm cloud}\sim 100 \Msun \pc^{-2}$)
typical of star-forming clouds in the Milky Way and Local Group
(e.g. \citealt{1987ApJ...319..730S,She08,Bol08,2009ApJ...699.1092H}), 
we therefore have $(M_{\rm atom}/M_{\rm mol})_{\rm GBC} < 1$ 
except for low metallicity
environments $Z'<0.2$. 
The regions of galaxies mapped in the \citet{Bigiel08} sample have 
metallicity $Z'\simgt 0.5$, so clouds with 
$N_{\rm H,cloud}\sim  10^{22}\cm^{-2}$ 
would be mostly molecular.
For clouds with this range of metallicity and column density such that 
$A_V\simgt 3$, most of 
the gas that is molecular (H$_2$ rather than \ion{H}{1}) would also be 
observable in CO \citep{2010ApJ...716.1191W,2010arXiv1003.1340G}.
Assuming that for moderate values of $\Sigma$, most of
the molecular gas is confined within GBCs, we also have 
$\Sigma_{\rm mol, diff}/\Sigma_{\rm mol,GBC}\ll 1$
(in galactic center regions where $\Sigma > 100\Msun \pc^{-2}$, the 
diffuse molecular fraction  may be larger).  Thus, in the regime for which 
$t_{\rm SF, CO}$ is observed to be approximately constant, 
$\siggbc/\Sigma_{\rm mol}\sim 1$ and $\Sigma_{\rm mol}\sim \Sigma_{\rm mol, CO}$,
and we can then take $\tsfgbc\approx t_{\rm SF, mol} \approx t_{\rm SF,CO}$.  
Note that because CO is optically 
thick in clouds with  $N_{\rm H,cloud}\sim  10^{22}\cm^{-2}$ 
and normal metallicity, observed CO emission in unresolved 
clouds  may in fact trace the atomic 
and ``dark gas'' portions of GBCs as well as the regions where CO is 
present, because these contribute to the 
gravitational potential and therefore the total CO linewidth.  To the extent 
that this is true (and provided there is minimal diffuse CO-emitting gas), 
$t_{\rm SF,CO}$ would be a direct measurement of $t_{\rm SF, GBC}$.

For the rest of this paper, we shall use $\tsfgbc\rightarrow
\tsf=const.$, and adopt the fiducial value of $2\times 10^9\yr$ based
on \citet{Bigiel08}.  Although the value we use for $\tsfgbc$ is
calibrated from observations in which star-forming clouds are
primarily molecular (and observable in CO lines), our basic approach
would remain unchanged for GBCs in different parameter regimes,
provided that a well-defined value of $\tsfgbc$ is known (from either
observations with appropriate corrections for atomic and dark gas, or
from theory).  From a theoretical point of view, the internal
dynamical properties of GBCs would be qualitatively similar whether
they are mostly molecular or a mixture of cold atomic and molecular
gas, but the value of $\tsfgbc$ would have to be adjusted to allow for
the dependence of star formation efficiency on chemical content (see
\S \ref{sec-consider}).

With the above assumptions, in equation (\ref{sigdiff-eq}) we now can
substitute $\sigdiff=\Sigma - \siggbc$ on the left-hand side, eliminate
$\Pth$ in favor of $\sigsfr$ on the right-hand side 
using equation (\ref{SFR-P-eq}) (or eqs. \ref{Wol-eq} -
\ref{zet-eq}), 
and set $\siggbc=\tsf \sigsfr$. The result is an
implicit expression for $\sigsfr$ in terms of $\Sigma$ and $\rhoext$, which
may be solved numerically.  The Appendix provides a derivation of the 
expression for $\sigsfr$, given in equations (\ref{x-eq}-\ref{SFR-eq}).
Closed-form analytic solutions may be obtained
in the limit of high and low values of the star formation rate. 
When $\sigsfr$ is high, the ISM is GBC-dominated 
and $\sigsfr \approx \Sigma/\tsf$.  When $\sigsfr$ is low,
the ISM is diffuse-dominated.  In this limit, $\sigdiff \rightarrow \Sigma$ and 
we may drop $\siggbc/\sigdiff$ in equation (\ref{pth-eq}), resulting in 
\begin{equation}
\frac{P_{\rm th,low}}{k}= \frac{  1700 \K \pcc}{\alpha} 
\left(\frac{\Sigma}{10 \Msun \pc^{-2}}\right)^2
\left\{
{1 +
\left[1 +  
50\fwt \alpha\frac{\left(\frac{\rhoext}{0.1\Msun\pc^{-3}  }\right)}
{\left(\frac{\Sigma}{10\Msun\pc^{-2} }\right)^2 }\right]^{1/2}}
\right\}.
\label{pth-dimen-eq}
\end{equation}
This result can then be inserted in equation (\ref{SFR-P-eq}) to
obtain a prediction for the dependence of $\sigsfr$ on $\rhoext$ and $\Sigma$ 
in low-density outer-disk regions.  An approximate form is
\begin{eqnarray}
\Sigma_{\rm SFR,low}&=& 3\times 10^{-10}\Msun\pc^{-2}\yr^{-1}\left(\frac{\Sigma}{10 \Msun \pc^{-2}}\right)\left[1+ 3 \left(\frac{Z_d'\Sigma}{10 \Msun\pc^{-2}} \right)^{0.4}  \right]
\times\nonumber \\
&  & \hskip 1cm\left[
\frac{ 2 }{\alpha  }\left(\frac{\Sigma}{10 \Msun \pc^{-2}}\right)
+ \left(\frac{50 \fwt }{\alpha  }  \right)^{1/2}
\left(\frac{\rhoext}{0.1\Msun\pc^{-3}  }\right)^{1/2}
\right]   
\label{asymp-SFR}
\end{eqnarray}
(see also eq. \ref{low-sig-lim} in the Appendix).
Note that the limit in equation (\ref{asymp-SFR}) applies in the 
Solar neighborhood, where the stellar+dark matter gravity (i.e. the 
term depending on $\rhoext$) accounts for 80\% of the total
contribution in second square bracket.
An approximate form over the whole range $\Sigma \simlt 100\Msun\pc^{-2}$ is
given by
\begin{equation}
\sigsfr\approx\left[ \frac{\tsf }{\Sigma } + \frac{1 }{\Sigma_{\rm SFR, low} }
\right]^{-1}
\label{SFR-approx}
\end{equation}
(see also eq. \ref{gbc-frac} in the Appendix).
In terms of the dependence on model parameters, at high surface density
$\sigsfr\propto \tsf^{-1}$, whereas at low surface density 
$\sigsfr\propto (\fwt/\alpha)^{1/2}(\Sigma_{\rm SFR,0}/P_{\rm th,0})$.

\subsection{Approach to equilibrium
\label{section-approach}
}

The above analysis yields a relation for the star formation rate in
equilibrium, but how would the self-consistent state of dynamical,
thermal, and star formation equilibrium be attained in a
real galaxy?  First, consider the timescales involved.  Vertical
dynamical equilibrium is reached on a timescale of a few times 
$\sim H/\sigma_z$, where
$\sigma_z$ is the vertical velocity dispersion. With $H\sim 100\pc$ and
$\sigma_z\sim 10\kms$, this dynamical equilibrium is typically reached
within a few times $10 \Myr$.  Thermal equilibrium of diffuse gas at a given
density is accomplished by a combination of heating and cooling; it is
very fast in dense gas, and for low density gas requires $\sim 5 \Myr$
\citep{Wol03}.  Separation of the diffuse gas into phases, which
involves dynamics and takes place via thermal instability, requires
$\sim 20-50 \Myr$ \citep{2004ApJ...601..905P,PO05}.  The timescale for
star formation to reach equilibrium depends on how GBCs are formed out
of the diffuse gas; recent simulations \citep{KO09a} suggest this may
take a several tens of $\Myr$ for conditions similar to the Solar
circle (but without spiral structure).  Thus, the slowest equilibrium
to be established is likely that of the star formation rate, which
depends on the relative proportions of diffuse and
gravitationally-bound gas.  Because GBCs are continually forming
and being destroyed within the disk (and material for a given GBC may be 
gathered horizontally from up to several $H$), at any time a given patch 
of the disk could be at a different point in this cycle.  In an observed 
galaxy, measuring the equilibrium properties (for given $\Sigma$ and $\rhoext$) 
would require averaging over a 
horizontal area large enough that the different states in the GBC
formation-destruction cycle are represented.

As a (highly idealized) example of how the system might evolve,
consider a region in which $\fdiff=\sigdiff/\Sigma$ is initially high
compared to the value in which all equilibrium conditions are
satisfied. On the one hand, a higher-than-equilibrium
$\fdiff$ implies a lower-than-equilibrium proportion of gas in
star-forming bound clouds, yielding a value for
$\sigsfr=(1-\fdiff)\Sigma/\tsf$ lower than the level when overall
equilibrium (including star formation equilibrium) obtains.  From
equations (\ref{Wol-eq}) and (\ref{G0-eq}), when the radiation field
is weak, the heating-cooling curve has a low values of $P_{\rm max,
  warm}$ and $P_{\rm min, cold}$ 
(and $P_{\rm two-phase}$).  
On the other hand, we
would expect that within a few tens of Myr, thermal, dynamical, and
phase equilibrium would be established with this (high) value of
$\sigdiff$.  From equation (\ref{pth-eq}), if $\fwt$ is constant, the
midplane pressure in the diffuse gas in this situation would be higher
than the value that would obtain when star formation equilibrium is
satisfied (it can be shown algebraically from eq. \ref{pth-eq} that
$\partial \Pth/\partial \fdiff>0$).  Thus, for this situation with
lower-than-equilibrium star-formation rate (but other processes in
equilibrium), the cooling equilibrium curve sits at a pressure lower
than that in overall equilibrium, while the midplane pressure is
higher than that in overall equilibrium.

\begin{figure}
\epsscale{.35}
%\epsscale{1.5}
\plotone{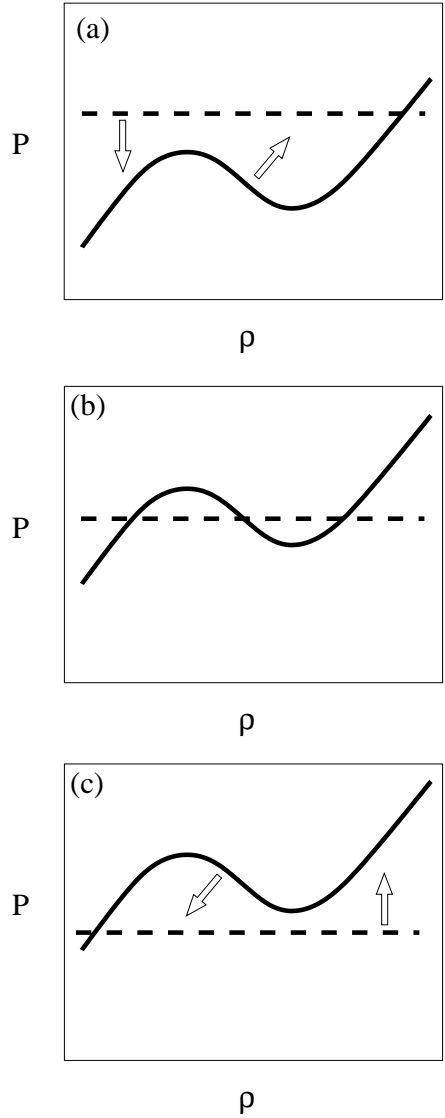}
\caption{Schematic of approach to overall equilibrium.  In (a), a 
higher-than-equilibrium value of the diffuse fraction $\fdiff$ makes 
the midplane pressure in the diffuse gas 
(dashed line) too high, while weak heating when 
the bound-cloud abundance $1-\fdiff$ 
is lower-than-equilibrium makes the thermal equilibrium
curve (solid line) lie at too-low pressure.  In (c), $\fdiff$ is 
lower-than-equilibrium, making the midplane pressure in 
the diffuse gas (dashed line) too low; at the same time, excess  
heating due to a high abundance of star-forming bound clouds 
situates the thermal equilibrium curve (solid line) too high.  
In (b), $\fdiff$ is
appropriate for simultaneous dynamical and thermal equilibrium with 
$\Pth$ at the midplane equal to 
$P_{\rm two-phase}$.
Arrows in (a) and (c) indicate the direction of evolution of the midplane 
pressure level and thermal equilibrium curve for $\fdiff$ to approach 
the equilibrium value.
\label{schematic-fig}}
\end{figure}

Figure (\ref{schematic-fig}a) shows an extreme case of 
higher-than-equilibrium $\fdiff$, 
in which the midplane pressure is higher than $P_{\rm max, warm}$.
In this situation, a portion of the warm gas would condense to make 
cold clouds, lowering $\fwt$ and reducing the midplane pressure.  
With a higher abundance of cold clouds, additional GBCs would form, 
lowering $\fdiff$, raising $\sigsfr$, and hence moving the
thermal equilibrium curve toward higher pressure.  As cold clouds are 
converted to GBCs, $\fwt$ could return to its equilibrium value.  But, 
with lower $\fdiff$ than in the initial situation, the midplane pressure would
be reduced.  The arrows in Fig. (\ref{schematic-fig}a) indicate how 
the midplane pressure and thermal equilibrium curve would evolve.
This process would continue until a self-consistent state of 
thermal, dynamical, and star formation equilibrium, with
$P_{\rm two-phase}$ 
comparable to the midplane pressure $\Pth$, is reached
(see Figure \ref{schematic-fig}b). (In fact, the midplane pressure
$\Pth$ only needs to lie between $P_{\rm min,cold}$ and $P_{\rm max,
  warm}$, which allows a range of self-consistent equilibria; we
discuss this issue below.)

As another example, we consider the opposite situation.  We imagine a
region with a bound cloud proportion $1-\fdiff$ and $\sigsfr/\Sigma$
above the self-consistent overall equilibrium values, such that heating
associated with the high star formation rate makes the
thermal equilibrium curve sit at high pressure (i.e. 
$P_{\rm two-phase}$ 
is high).  
With low $\fdiff$,
$\sigdiff=\fdiff \Sigma$ and the midplane diffuse-gas pressure will be
low compared to the overall equilibrium (assuming constant $\fwt$).
Figure (\ref{schematic-fig}c) illustrates an extreme of this situation,
in which the heating rate is so high that cold atomic clouds would
secularly heat and expand to increase the proportion of warm gas.  The
energy input to bound clouds associated with the high star formation
rate would destroy GBCs at a faster rate than they could form,
increasing $\fdiff$ and lowering $\sigsfr$ until a self-consistent
equilibrium state (Figure \ref{schematic-fig}b) is reached.  The
arrows indicate the evolution required move to a state in which
thermal, dynamical, and star formation equilibrium are all satisfied.

We note that for a given $\Sigma$ and 
$\fdiff$, which fixes $\siggbc$ and therefore $\sigsfr$, $G_0'$ and 
$P_{\rm two-phase}$ 
for the cooling curve, the ISM conditions
would still be consistent with a two-phase medium for all values of
the midplane pressure $\Pth$ between $P_{\rm min,cold}$ and $P_{\rm
  max,warm}$. Since $\Pth$ varies approximately 
$\propto(\fwt/\alpha)^{1/2}= \fwt(\fwt + v_t^2/c_w^2)^{-1/2}$
(see eq. \ref{approx-P}), even with $\fwt \rightarrow 1$ the
midplane pressure would increase by less than a factor of two (remaining
$<P_{\rm max,warm}$)
 compared to the solution assuming $\fwt=0.5$.  A value $\fwt\sim 0.2$ would 
reduce $\Pth$ by a
factor $\sim 2$ to approach $P_{\rm min,cold}$.  

Although a range of $\fwt$
is thermodynamically permitted, self-consistent numerical simulations
are required to ascertain how wide a range of $\fwt$ can actually be
realized, since a number of physical processes enter in setting $\fwt$.  
For a given $f_{\rm diff}$ and $\Sigma$ and hence a given heating rate 
(and thus a fixed thermal equilibrium curve), solutions
with a larger fraction of the diffuse mass in the cold phase have lower
total energy; thermal instability and condensation of warm gas into cold clouds 
thus tends to drive the system toward
lower $\fwt$. This is limited, however, by turbulent mixing and thermal 
conduction, which tend to raise $\fwt$ and also produce 
out-of-equilibrium gas (see e.g. \citealt{PO05,PO07,2005A&A...433....1A}).
Another consideration is that if the value of $\fwt$ is 
too large (or too small), GBCs would not 
form sufficiently rapidly (or would form too rapidly) to maintain an 
equilibrium population with a total surface density 
$\siggbc=(1-\fdiff)\Sigma$. 
Yet another issue is that the turbulent velocity dispersion $v_t$ 
in the diffuse gas is likely maintained largely by energy inputs from 
star formation; with a lower star formation rate and $v_t$, the value 
of $\fwt$ would also have to decrease in order to maintain vertical 
dynamical equilibrium.  With self-consistent numerical simulations, it will
be possible to assess whether $\fwt$ and 
$\Pth/P_{\rm two-phase}$
secularly depend on $\Sigma$ and $\rhoext$.  For present purposes, we proceed
under the (observationally-motivated) assumptions that 
$\Pth \approx P_{\rm two-phase}$ 
and $\fwt \sim 0.5-1$.  The former assumption is also
motivated by theory, as discussed in \S \ref{section-energy}.

\subsection{Sample solution for an idealized galaxy
\label{section-ideal}
}

\begin{figure}
\epsscale{.8}
\plotone{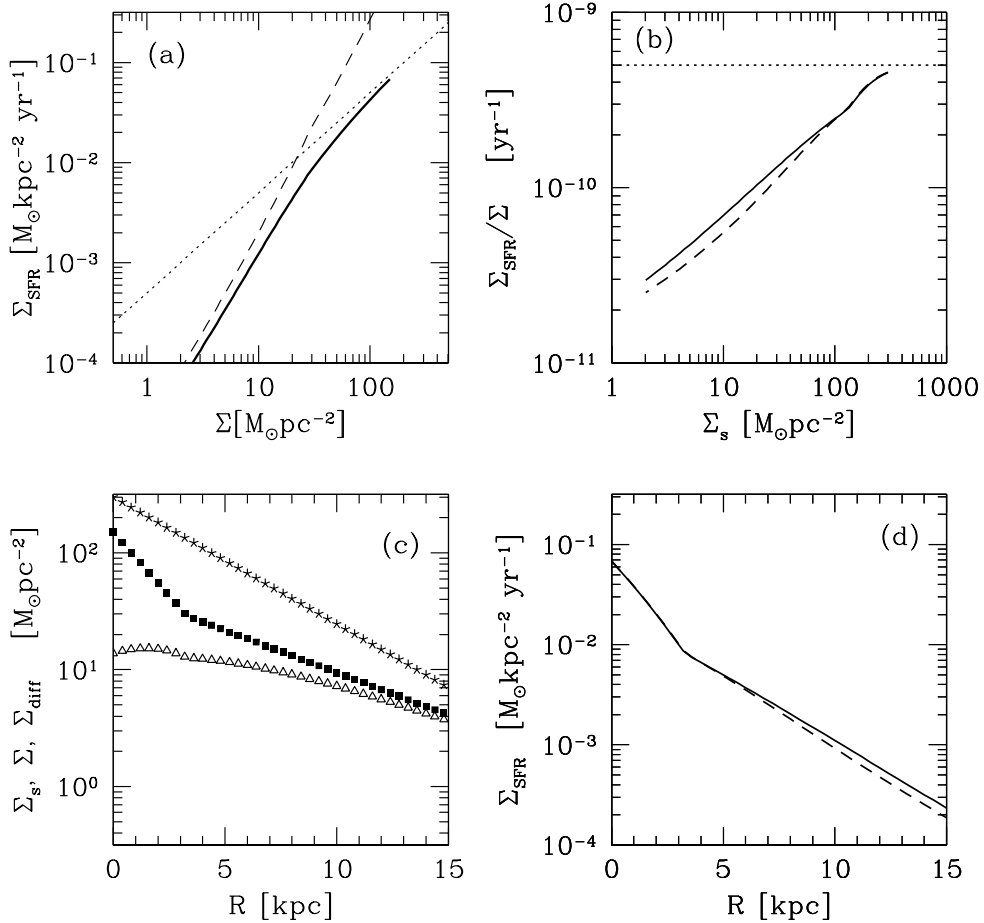}
\caption{Sample solution of the self-consistent star formation rate 
for an
idealized galactic model:
{(a)}
shows $\sigsfr$ as a function of $\Sigma$ ({\it solid}), together 
with the maximum possible rate $\Sigma/\tsf$ ({\it dotted}), and the 
outer-disk limiting solution given in equation 
(\ref{asymp-SFR}) ({\it dashed});
{(b) } 
shows the ratio $\sigsfr/\Sigma$ as a function of $\Sigma_s$,
% ({\it solid}), 
together with the maximum value $1/\tsf$ ({\it dotted});
{ (c)} shows the input stellar surface
density $\Sigma_s$ 
({\it stars}) and input total gas surface density $\Sigma$ ({\it squares}) 
as a function of radius, together with the solution for $\sigdiff$ 
({\it triangles}), all as a function of galactic radius $R$; 
 { (d)}
shows the solution for $\sigsfr$ as a function of $R$.  
In (b) and (d), in addition to the case of  constant stellar disk thickness 
({\it solid}), 
results for a flaring disk $H_s\propto 1/\Sigma_s$ are shown as {\it dashed}
lines.
\label{fig-examp}}
\end{figure}

As an example of 
the predictions from this model, we consider an 
idealized disk galaxy in which the stellar surface density obeys an exponential,
$\Sigma_s (R)= \Sigma_s(0) \exp(-R/R_s)$; the total gaseous surface
density obeys a two part exponential, $\Sigma (R) = \Sigma_{g1}(0)
\exp(-R/R_{g1})$ for the inner disk, and similarly for the outer disk with 
$\Sigma_{g2}$ and $R_{g2}$; 
and the rotation velocity contribution 
from the dark-matter halo is $V(R)=V_c [1-\exp(-R/R_h)]$. 
Setting  model parameters equal to $\Sigma_s=300 \Msun\pc^{-2}$, 
$\Sigma_{g1}=150 \Msun\pc^{-2}$,
$\Sigma_{g2}=50 \Msun\pc^{-2}$,  
$R_s=4\kpc$, $R_{g1}=2\kpc$, $R_{g2}=6\kpc$,
$V_c=200\kms$, and $R_h=1\kpc$,
Figure (\ref{fig-examp}) shows the adopted 
surface density profiles as well as the 
solution for $\sigsfr$ and $\sigdiff$.  

For this example, results shown
in Fig. \ref{fig-examp} (a) and (c)   adopt
a constant stellar disk thickness $H_s$ such that 
$\rho_s=\Sigma_s/(0.54 R_s)$ following 
\citet{Ler08}.\footnote{Note that this choice of coefficient 
may underestimate the midplane density somewhat.  For 
the local Milky Way, which has $\rho_s=0.04\Msun \pc^{-3}$ 
\citep{Hol00} and $\Sigma_s=42 \Msun\pc^{-2}$ \citep{2004MNRAS.352..440H} 
from dynamical mass models, a scale length of $R_s=2.6\kpc$
\citep{2008ApJ...673..864J} would yield 
$\Sigma_s/(0.54 R_s)=0.03\Msun  \pc^{-3}$.} 
The solution is 
not very sensitive to this choice, however; to show this, in 
Fig. \ref{fig-examp} (b) and (d), we compare results for both 
constant-thickness stellar disk $H_s =const.\equiv H_{S,0}$
and flaring stellar disk $H_s (R)=H_{S,0}\Sigma_s(R_s)/\Sigma_s(R)$.
In general, from equation
(\ref{asymp-SFR}), $\sigsfr/\Sigma \propto \rhoext^{1/2}$ in the outer
disk.  Thus, if the stellar disk dominates, 
$\sigsfr/\Sigma \propto \rho_s^{1/2} \propto (\Sigma_s/H_s)^{1/2}$, which 
yields either $\sigsfr/\Sigma \propto \Sigma_s^{1/2}$
if $H_s=const.$, or 
$\sigsfr/\Sigma \propto \Sigma_s$ if the outer disk is flaring
with $v_{z,s}=const.$ and $H_s=v_{z,s}^2/(\pi G \Sigma_s)$.  
If the dark 
matter density dominates the stellar density in the outer disk, then since 
$\rhodm\propto (V_c/R)^2$, constant $V_c$ would imply 
$\sigsfr/\Sigma \propto R^{-1}$ in the outer disk.

The example shown illustrates several characteristic features 
that are in accord with the recent observational results of \citet{Bigiel08} 
and \citet{Ler08}.
First, it is evident that star formation does not have a sharp cutoff in
the outer disk, but instead the rate gradually declines with $R$ and 
$\Sigma_s$ (see also the GALEX observations of 
\citealt{2007ApJS..173..524B}).  Second, there 
are two regimes evident for $\sigsfr$ and $\sigdiff$ vs. $\Sigma$:
a high-surface-density regime in
which the gas is mostly in self-gravitating clouds and the limiting solution
$\sigsfr = \Sigma/\tsf$ is approached, and a low-surface density regime in
which $\sigsfr$ has a steeper dependence.  From equation 
(\ref{asymp-SFR}), the predicted limiting behavior in outer disks 
(or dwarfs) is  
$\sigsfr \propto \Sigma \rhoext^{1/2}$, so that a dropoff in the stellar and
dark-matter density with $R$ steeper than we have adopted, or a decline in
$\Sigma$ with $R$ 
shallower than we have adopted (for a given $\rhoext(R)$), 
would yield a steeper dependence 
of $\sigsfr$ on $\Sigma$ in the outer disk.\footnote{Note that in the 
limit of $\Sigma$ constant, the Schmidt-law slope would be infinite,
because each value of $\rhoext(R)$ would yield a different value of 
$\Sigma_{\rm SFR}$.}
That is, there is no single 
universal slope predicted for the Schmidt law in outer disks, since 
$\rhoext^{1/2}$ need not obey a power law $\propto \Sigma^p$.
In individual galaxies, observations in fact show a
range of behaviors for $\sigsfr$ vs. $\Sigma$ 
ranging from steep declines to slow taper at low $\Sigma$, with 
an upper envelope $\sigsfr = \Sigma/\tsf$ at high $\Sigma$.

An interesting feature evident in Figure (\ref{fig-examp})
is that, even while the total gas surface density rises sharply in the inner
disk, $\sigdiff$ does not.  It is straightforward to set an
upper limit on $\sigdiff$: from equation (\ref{sigdiff-eq}), 
$\sigdiff < \alpha \Pth/(\pi G \siggbc)$; using $\siggbc=\sigsfr \tsf$ and
equation (\ref{SFR-P-eq}),
\begin{equation}
\sigdiff< 36 \Msun \pc^{-2} \frac{\alpha}
{1+ 3 \left(\frac{Z_d'\Sigma}{\Sigma_0} \right)^{0.4}  }.
\label{Siglim-eq}
\end{equation}
The physical reason for this limit is that in regions of very active 
star formation, the enhancement in UV heating -- which tends to produce 
diffuse atomic gas -- coincides with stronger compression
of the gas by the gravity of the massive, star-forming clouds.  Under 
high-pressure conditions, the cooling rate increases; if the
diffuse gas density is too high, 
cooling will exceed heating, and diffuse gas will be driven into 
the self-gravitating component.
Since $\alpha \sim 2-10$ 
(see eq. \ref{alpha_def})
with a value $\sim 3-5$ most likely,
the formal limit in equation (\ref{Siglim-eq}) 
starts to become constraining only in inner galaxies where 
$\Sigma/\Sigma_0 \gg 1$.
As discussed in the Appendix (see eq. \ref{sigdiff-lim}), in fact the 
terms from the (variable) stellar and dark matter gravity
in equation (\ref{sigdiff-eq}) reduce $\sigdiff$ by another factor $\sim 2$ 
relative to the upper limit in equation (\ref{Siglim-eq}).

In adopting $G_0'=\sigsfr/\Sigma_{\rm SFR,0}$ for equation (\ref{Wol-eq}), 
we have neglected optical depth effects. 
In regions of high $\sigdiff$ and/or high $Z'$ such that the optical depth is
high, the mean (scaled) UV intensity $G_0'$ 
within the diffuse gas is reduced 
(by a factor $\sim [1-E_2(\tau_\perp/2)]/\tau_\perp$; this varies 
$\propto 1/\tau_\perp$ for large $\tau_\perp$), 
consequently lowering $\Pth$ and, via equation 
(\ref{sigdiff-eq}), $\sigdiff$ itself.  
Thus, the current simple theory overestimates $\sigdiff$ in the central parts
of galaxies, where $\tau_\perp$ becomes large.
We further 
note (see \S \ref{sec-disc}) that
the value of $\sigdiff$ is not equivalent to the surface density of 
atomic gas, since GBCs can have atomic envelopes (or even be 
mostly atomic at low $Z$) and since diffuse gas 
can be molecular (at high $\Sigma Z$).

\subsection{Additional considerations
\label{sec-consider}
}

Finally, we remark on a few additional points related to assumptions behind and 
application of the theory presented above:

\medskip
%\leftline
{\it 1. Correction for hot gas -- } 
The formulae we have developed assume that on the scales over
which horizontal averages are taken, the area filling factor of hot
gas $f_{\rm hot}$ 
is negligible.  In comparing to observations, if the fraction of
the area in a beam that is primarily warm+cold gas is $1-f_{\rm hot}$, then the
true surface density of warm+cold gas, for the purpose of computing
dynamical and thermal equilibrium, is $\Sigma=\Sigma_{\rm obs}/(1-f_{\rm hot})$,
and similarly for $\sigdiff$ and $\siggbc$.  For given $\Sigma_{\rm obs}$,
increasing $f_{\rm hot}$ would tend to increase the gaseous gravity terms in
equation (\ref{pth-eq}) relative to the stellar/dark matter term, and
would raise the overall value of the midplane pressure.

\medskip
%\leftline
{\it 2. The parameter $\fwt$ --} 
In the foregoing analysis, we have treated $\fwt$ 
as an exogenous parameter that -- based on 
observations -- does not vary strongly within a given galaxy or from one
galaxy to another.  In reality, the relative proportions of cold and warm 
atomic gas must be self-consistently determined by considering the overall
cycle of gas among phases, a highly complex problem.  By considering a simple
limiting case, however, it is possible to see why $\fwt$ might have 
only moderate variation. 
We suppose that the primary way GBCs
form is through self-gravitating collection of the cold diffuse atomic  
medium, at a rate per unit area $\Sigma_c/t_{g,c}$ where 
$t_{g,c}\sim(G \Sigma_c/(\sqrt{2\pi}H_c))^{-1/2}$ is the self-gravitational 
timescale for the distribution of cold atomic clouds, which has total surface
density $\Sigma_c\approx(1-\fwt)\sigdiff$ and vertical thickness 
$H_c \sim v_t/\sqrt{4 \pi G \rhoext}$ (in the case where stars and dark 
matter dominate the vertical gravity).  We also suppose that 
GBCs
have mean lifetimes 
$t_{\rm GBC}$ ($\sim 20 \Myr$; e.g. \citealt{2007prpl.conf...81B}), 
such that the rate of destruction of 
the gravitationally-bound component per unit area is 
$\siggbc/t_{\rm GBC}=\sigsfr \tsf/t_{\rm GBC}$.  Equating the formation and 
destruction rates yields
\begin{equation}
1-\fwt= \left(\frac{\tsf}{t_{\rm GBC}}  \right)^{2/3}
\left(\frac{ v_t^2 }{2G^3  }  \right)^{1/6}
\frac{\sigsfr^{2/3}}{\sigdiff \rhoext^{1/6}  }.
\label{GBC-form}
\end{equation}
Substituting in Solar-neighborhood parameters on the
right-hand side, this yields $\fwt\sim 0.6$, in agreement with local
observed 
estimates.  Next, if we consider the diffuse-dominated 
$\sigdiff \approx \Sigma$ limit of 
equation (\ref{asymp-SFR}) (applicability of 
this limit includes the Solar neighborhood) and
substitute in $\sigsfr \propto \Sigma \sqrt{\fwt \rhoext}$ on the
right-hand side of equation (\ref{GBC-form}), 
we see that $(1-\fwt)/\fwt^{1/3} \propto
(\rhoext/\Sigma^2)^{1/6}$.  The weak dependence on both $\rhoext$ and
$\Sigma^2$, and the fact that only their ratio appears so that
variations will be partially compensated, implies that
$\fwt$ would indeed be expected to vary only modestly, at least in outer disks.

Of course, the above is an (oversimplified) description of only one of the 
possible ways in which GBCs might form.  Understanding how 
$\fwt$ depends on the fundamental environmental properties of a galaxy 
($\Sigma$, $\rhoext$, metallicity) will require numerical simulations that 
follow a wide range of processes, including realistic treatment of turbulence
(which can alter $\fwt$ via small-scale local mixing, and can also collect
diffuse gas into GBCs if large-scale flows with long durations are present).

\medskip
%\leftline
{\it 3. Galactic and metagalactic radiation -- } 
In adopting $G_0'=\sigsfr/\Sigma_{SFR,0}$, we have assumed that the contribution
from metagalactic UV is small compared to the locally-generated intensity
(and neglected the slowly-varying dependence on optical depth).
\citet{2002ApJS..143..419S} estimate that the ratio of metagalactic to 
locally-generated UV in the Solar neighborhood is 0.0024.  The star formation
rate could therefore decline to $\sim 0.002$ times the local value, or 
$6 \times 10^{-6}\Msun \kpc^{-2} \yr^{-1}$, before the metagalactic UV 
becomes important; this occurs only in the far outer regions of disks.  In
this regime, instead of using $G_0'=\sigsfr/\Sigma_{SFR,0}$ in equation 
(\ref{Wol-eq}), a constant value $\sim 0.002$ would be substituted.  
Beyond this point, 
$P_{\rm two-phase}$ 
would be essentially constant from equation 
(\ref{Wol-eq}), and the midplane pressure (given by eq. \ref{pth-dimen-eq}) 
would fall below $P_{\rm min,cold}$ (as in Fig. \ref{schematic-fig}c).  
Beyond this point, the diffuse gas would 
be essentially all warm, and star formation could only occur to the extent 
that gas is externally compressed, e.g. by spiral density waves.
In practice, since in our theory 
$\sigsfr \propto \Sigma \sqrt{\rhoext}\propto \Sigma V_c/R$ 
in outer disks if dark matter dominates the gravity, the gas may 
reach sufficiently low surface density ($\sim 0.1 \Msun \pc^{-2}$; see 
\citealt{2002ApJS..143..419S}) that it could be ionized
by metagalactic X-rays before the point where
$\Sigma\simlt 0.002 \Msun \pc^{-2}R/\kpc$ is reached.
In addition to true metagalactic radiation, 
the contribution to the UV intensity originating nonlocally within 
a galaxy climbs as the local optical depth drops. At very low optical 
depths where
$J_{\rm FUV} \propto \sigsfr \ln(1/\tau_\perp)$, the star formation rate
would be reduced below that in equation (\ref{asymp-SFR}) by a factor 
$\sim 1/\ln(1/\tau_\perp)$; this varies $\propto 1/R$ if the gas surface 
density obeys an exponential.  In far outer galaxy regions, the radiation 
from the inner galaxy may exceed that produced by local star formation, 
expecially if the disk is strongly flaring.

\medskip
%\leftline
{\it 4. Properties of GBCs -- } 
In this work, we have not focused on the details of individual GBCs,
beyond making the simplifying assumption that the gas in these
structures forms stars on a timescale $\tsf$ that does not
significantly vary from one cloud to another.  At a fundamental level
(see \citealt{MO07}), the star formation timescale within a GBC is
expected to depend on the mean density (which sets the mean gravitational 
free-fall time $t_{\rm ff}$) and on the amplitude of
turbulence and strength of the magnetic field (since these properties
determine how gas is further compressed and rarefied). 

If the star formation
rate in a GBC is defined to be 
$\epsff M_{\rm cloud}/t_{\rm ff}$, then
$\tsfgbc=t_{\rm ff}/\epsff$, where 
$t_{\rm ff}=(\pi M_{\rm cloud}/\Sigma_{\rm cloud}^3)^{1/4}(8 G)^{-1/2}$ in terms of
the cloud's mass and mean surface density. \citet{KM05} have argued that,
due to the lognormal form of the density PDF in turbulent clouds, 
$\epsff$ will depend only weakly on a cloud's 
internal turbulent Mach number, which itself varies as 
$v_{\rm turb}/v_{\rm th}\propto (M_{\rm cloud}\Sigma_{\rm cloud})^{1/4} 
T_{\rm cloud}^{-1/2}$.  
As a consquence,
$\tsfgbc$  is not expected to vary very strongly with a cloud's properties; 
e.g. \citet{KM05} propose a scaling which yields
$\tsfgbc \propto M_{\rm cloud}^{1/3} 
\Sigma_{\rm cloud}^{-2/3} T_{\rm cloud}^{-1/6}$.  As noted above, GBCs
are composed of both molecular (shielded) and atomic (shielding) gas.
Because the atomic gas has temperature somewhat higher ($\simlt 100\K$) 
than that of the 
molecular gas ($\simlt 10\K$), the star formation efficiency may vary within a 
given GBC, as well as varying from one cloud to another.  
Since the temperature is determined by cooling, it depends on whether
carbon is mostly atomic or in CO, with the latter holding in the more-shielded
parts of clouds \citep{2010ApJ...716.1191W}.
In a more refined theory, these intra- and inter-cloud variations 
could be taken into account in determining a mean value of 
$\tsfgbc$ (for an assumed cloud mass function); here, 
we have simply adopted a single constant value, $\tsf$.

It is worth emphasizing again that the GBC component in our model is not 
equivalent to the molecular component observed in galaxies.  An individual
GBC is composed of a mixture of molecular gas and cold atomic gas 
that depends on shielding, and could be primarily atomic at sufficiently 
low metallicity.  For spherical clouds, 
$(M_{\rm atom}/M_{\rm mol})_{\rm cloud}\approx [Z'^{0.8}
  (N_{\rm H,cloud}/1.8\times 10^{21}\cm^{-2})-0.7]^{-1}$ 
\citep{KMT09,2010ApJ...709..308M}. 
Although the chemistry of GBCs depends strongly on $Z$, the temperature of
the cold gas is relatively insensitive to metallicity at high density
$n_H\simgt 100 \pcc$ (e.g. \citealt{Wol95}).
Thus, the internal dynamics of primarily-atomic GBCs -- including the 
processes that 
determine the internal star formation efficiency -- are expected to be 
similar to those in primarily-molecular GBCs, provided that 
their gravitational potentials and internal velocity dispersions 
are similar so that $v_{\rm turb}\gg v_{\rm th, cold}$.

We note that \citet{2009ApJ...699..850K} have developed a model for
galactic star formation rates under an alternative set of assumptions.
In their model, interstellar gas is assumed to be gathered into
complexes with mean surface densities $\Sigma_{\rm complex}\sim 5
\Sigma$, where $\Sigma$ is averaged over $\sim\kpc$ scales. The
fraction of mass within complexes that participates in star formation
is determined by shielding. The molecular gas is assumed to be in GMCs
with a surface density equal to the value observed in local galaxies,
$\Sigma_{\rm cloud}= 85 \Msun \pc^{-2}$, provided this exceeds the mean gas
surface density.  Stars form in the GMCs at a rate determined by the
\citet{KM05} theory.  In this model, $\sigsfr$ becomes a steep
function of $\Sigma$ when complexes become primarily atomic, for ISM
surface density 
$\Sigma \simlt (20/Z')(\Sigma/\Sigma_{\rm complex}) \Msun \pc^{-2}$.  
The \citet{2009ApJ...699..850K} model makes fewer assumptions
and extends to higher-$\Sigma$ conditions than the model discussed
here. However, while it is successful in describing the average
properties of star-forming galaxies, it is substantially less accurate
than the present model in describing the star formation in individual
galaxies, which is discussed below.

\section{Comparison to Observations
\label{sec-obs}
}

The formulae derived above yield predictions for $\sigsfr$ as a
function of galactic gas and stellar properties, and can be compared
with observations.  Here, we compare with a recent survey of spiral
galaxies, for which the gas, stellar, and star formation content is
described in detail in \citet{Ler08}.  The observed measurements
include molecular and atomic gaseous surface densities, $\Sigma_{\rm mol}$
and $\Sigma_{\rm atom}$ (based on CO $J=2-1$ and $1-0$ maps and on HI 
21 cm maps, respectively, and corrected for Helium), 
rotation curves $V_c(R)$, stellar surface
densities $\Sigma_s$ (based on 3.6 $\mu$m {\it Spitzer} maps), and
star formation surface densities $\sigsfr$ (based on FUV maps from
{\it GALEX} and 24 $\mu$m {\it Spitzer} maps).  The surveyed regions
include both molecule-dominated and atomic-dominated areas, extending
to $\sim 1.2 r_{25}$.

Here, we estimate stellar densities in two ways, taking the disk scale 
height $H_s=const.$ so that
$\rho_s(R)=\Sigma_s(R)/(0.54 R_s)$ following \citet{Ler08}; and taking
$H_s\propto 1/\Sigma_s$ (i.e. a flared disk) so that
$\rho_s(R)=\Sigma_s^2(R)/[0.54 R_s \Sigma_s(R_s)]$, where $R_s$ is the 
fitted exponential scale length of the stellar disk.  We estimate
dark matter densities using observed rotation curves, as
$\rhodm(R)= (V_c^2 - V_{c,s}^2)/(4\pi G R^2)$, where 
$V_{c,s}^2/R$ is the correction for the contribution to the radial 
acceleration from the stellar disk 
(this correction is $\simlt 50\%$ in the outer disk, where the 
contribution to vertical gravity from dark matter becomes significant).
Given $\rho_s$, $\rhodm$, and total gas surface 
density $\Sigma=\Sigma_{\rm atom} + \Sigma_{\rm mol}$ 
at each radius $R$, 
we numerically solve the equations developed in 
\S 2 to obtain predictions for  $\sigdiff$ and $\sigsfr$ 
(see the Appendix for details).
For the initial
comparisons presented here, we use annular averages of the data sets in 
each galaxy; ``pointwise'' comparisons we have 
made at map resolutions of 800 pc show similar results, in terms of the
mean values and the scatter in the observations and model predictions.

\begin{figure}
\epsscale{1.}
\plotone{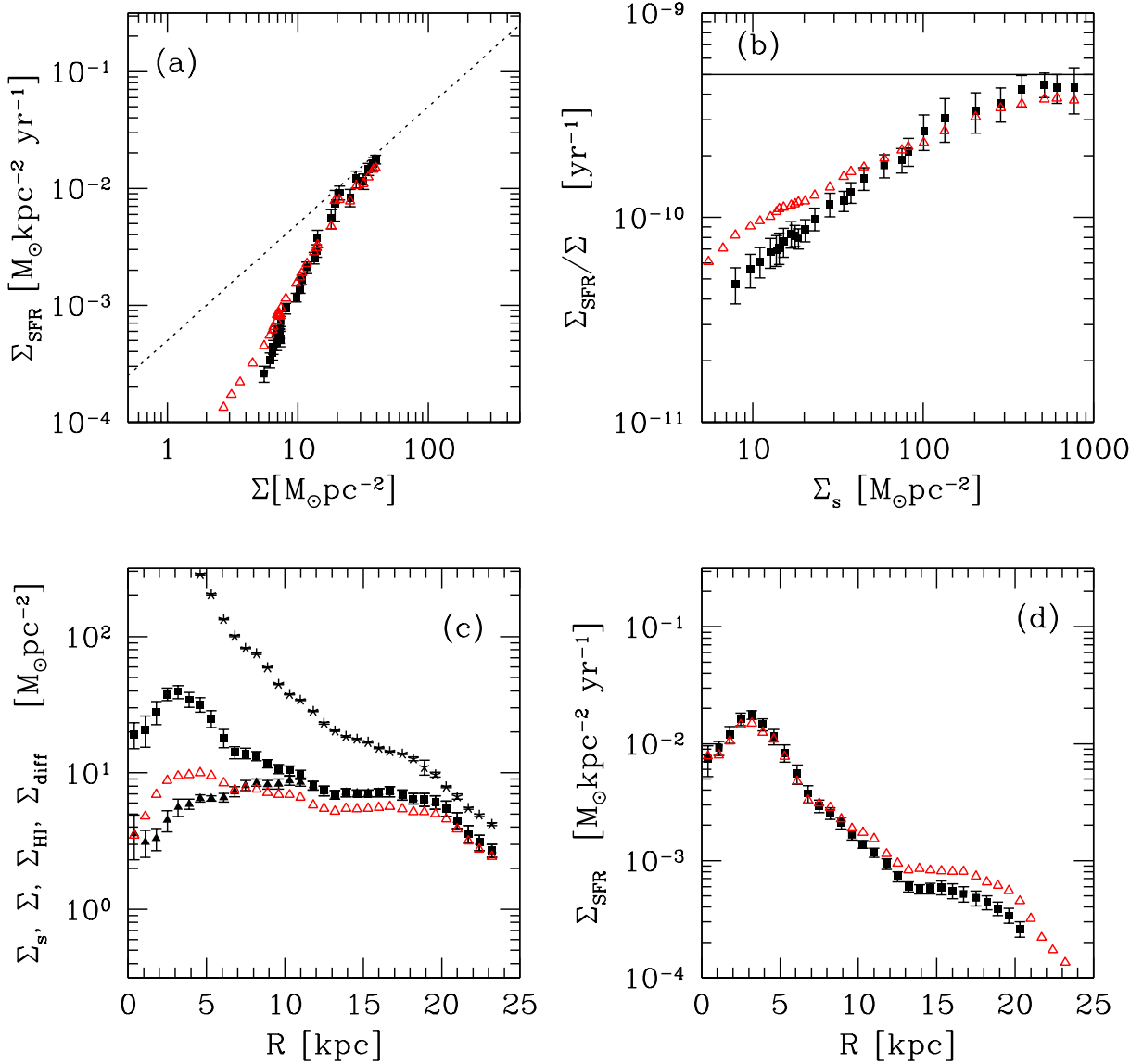}
\caption{Comparison between observed annular averages and model prediction  
for the
galaxy NGC 7331.  In (a), (b), and (d), filled squares with error bars 
show the observed values
of $\sigsfr$ or $\sigsfr/\Sigma$, where $\Sigma$ is the total gas surface 
density; open triangles ({\it red}) show the predictions from the 
theory of \S 2.  In (c), stars show the observed stellar surface density 
$\Sigma_s$, squares show total observed $\Sigma$, filled triangles 
show the observed atomic gas surface density, and open triangles ({\it red}) 
show the model prediction for the diffuse-gas surface density 
$\sigdiff$.  For the observations, error bars indicate scatter in the values
within each azimuthal ring (systematic errors are larger).
For the model prediction, 
$H_s=const.$ is adopted for the stellar scale height.
}
\label{NGC7331fig-const}
\end{figure}

\begin{figure}
\epsscale{1.}
\plotone{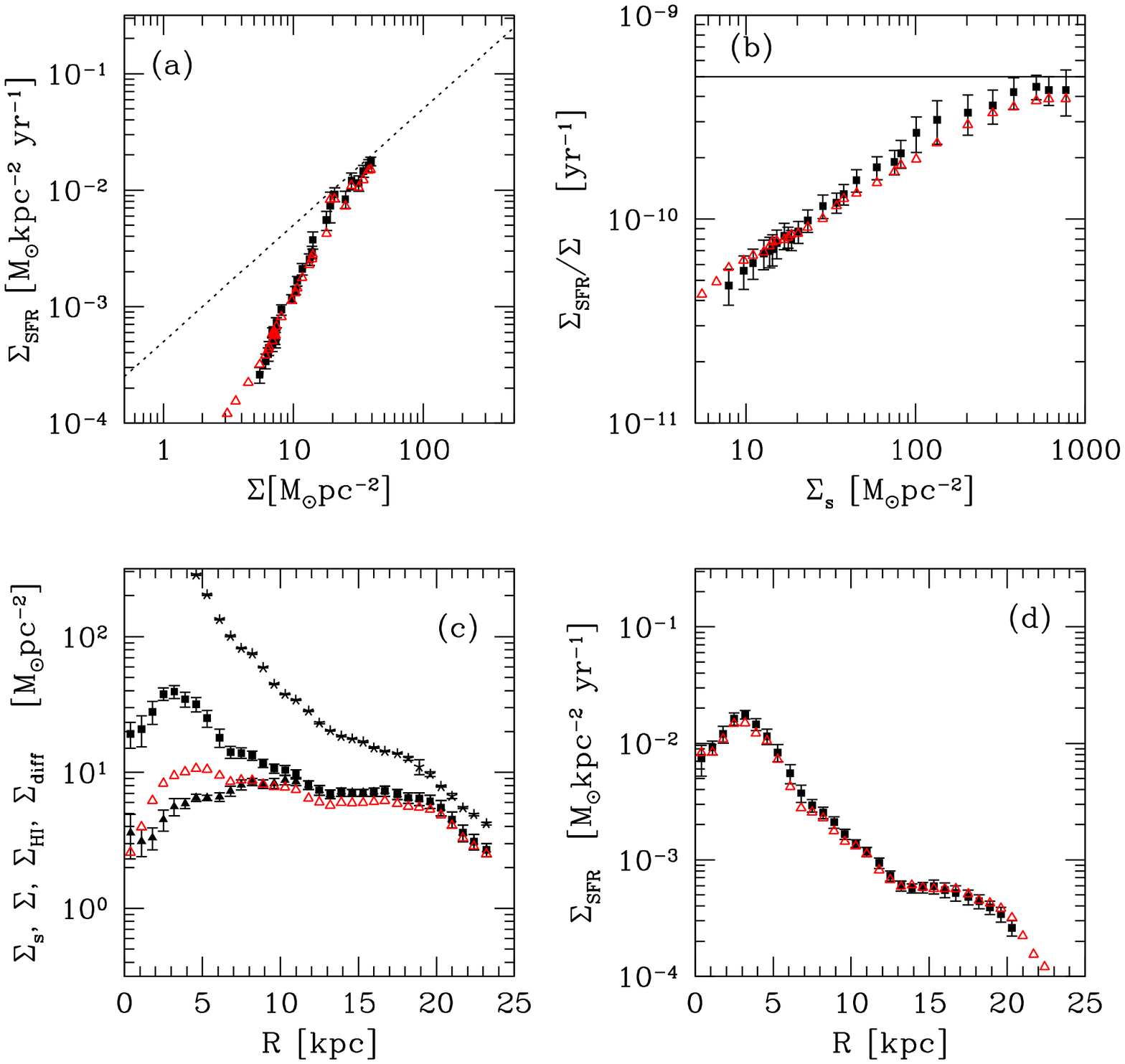}
\caption{Same as in Fig. \ref{NGC7331fig-const}, except a flaring stellar 
disk with $H_s\propto 1/\Sigma_s$ is adopted.
}
\label{NGC7331fig-flare}
\end{figure}

For the galaxy NGC 7331, Figures \ref{NGC7331fig-const} and
\ref{NGC7331fig-flare} present results for model predictions compared
to the observations, for flat and flared stellar disk cases 
$H_s=const.$ and $H_s\propto 1/\Sigma_s$, 
respectively.  For this galaxy, we have used the
fitted metallicity profile of \citet{1999ApJ...516...62D} and assumed
$Z_d'=Z_g'$, for equation (\ref{SFR-P-eq}).  We have adopted the fiducial
values $\alpha=5$ and $\fwt=0.5$ for the diffuse-ISM parameters, 
as discussed in \S 2. The values of $\sigsfr$ and
$\sigsfr/\Sigma$ are shown as functions of $R$, $\Sigma$, and
$\Sigma_s$.  Also shown are the input profiles of $\Sigma_s$,
$\Sigma=\Sigma_{\rm atom} + \Sigma_{\rm mol}$, and $\Sigma_{\rm atom}$, 
together with
the predicted $\sigdiff$.  Evidently, the model provides a remarkably
good prediction for $\sigsfr$, with a slightly better match for the
flared-disk case ($\simlt 20\%$ differences) than for the 
flat-disk case ($\simlt 50\%$ differences). 
In particular, the prediction follows the observation quite
well in the atomic-dominated regime (see Fig. \ref{NGC7331fig-flare}), outside
of $\approx 7\kpc$. We note that with slight adjustments of $\fwt/\alpha$,
the flaring of the stellar disk, or the dark matter density compared to 
the standard parameters and prescriptions, even closer agreement between 
predicted and observed $\sigsfr$ can be obtained.  Figures 
\ref{NGC7331fig-const} and \ref{NGC7331fig-flare} show that the model 
result for $\sigdiff$ exceeds $\Sigma_{\rm HI}$ in the inner region of
the galaxy. While some of the diffuse gas in galactic center regions could
in fact be molecular, we note (see \S \ref{section-ideal}) 
that our neglect of radiative 
transfer in estimating $J_{\rm FUV}$ makes $\sigdiff$ increasingly
inaccurate in regions of high $\Sigma$.  
This does not affect the predicted value of $\sigsfr$, however, 
because $\sigdiff \ll \Sigma$.

\begin{figure}
\epsscale{1.}
\plotone{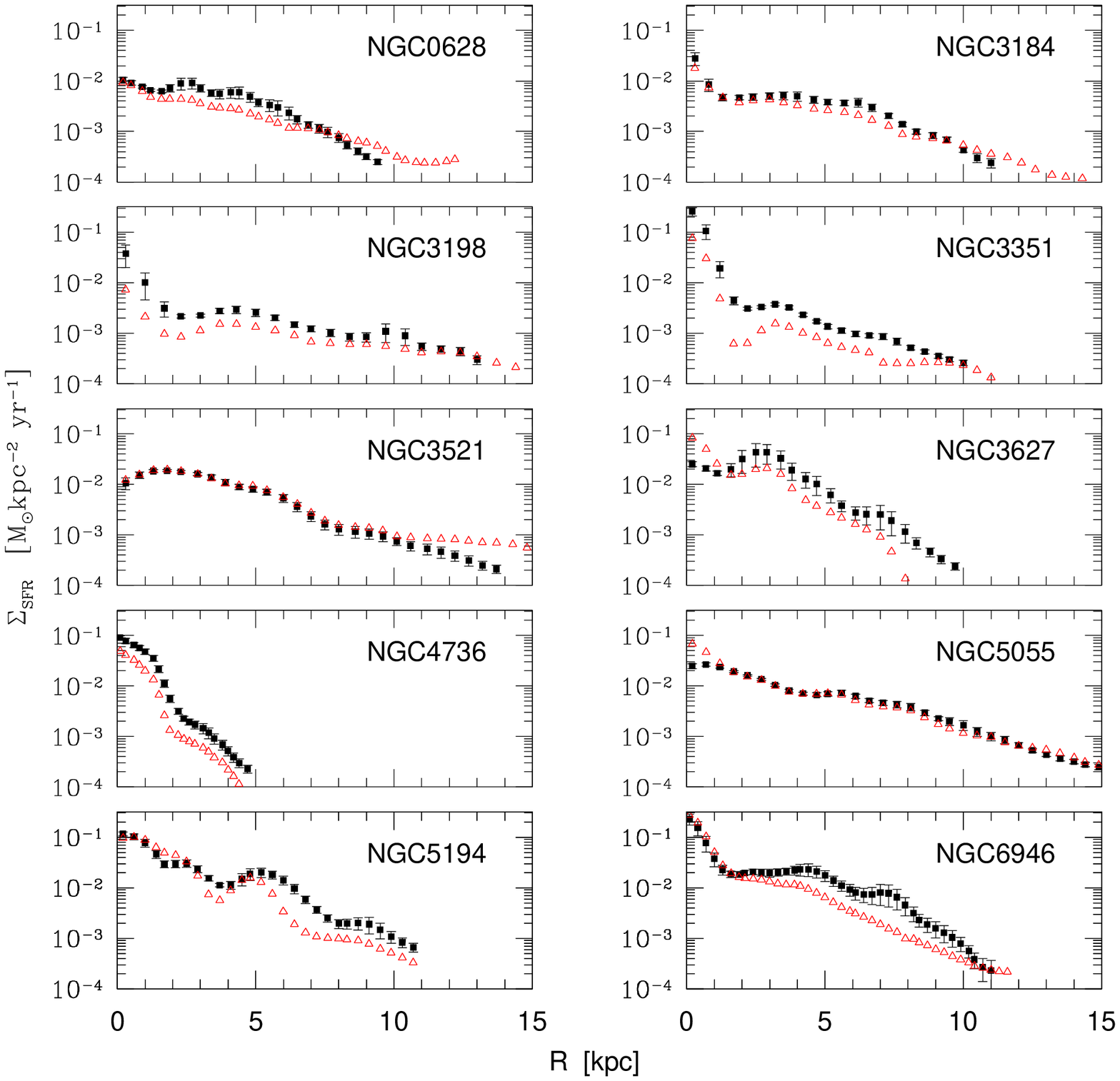}
\caption{Comparison between observed annular averages of $\sigsfr$ 
({\it black squares}) and 
model prediction ({\it red triangles}), for a set of spiral galaxies.  
For the model, $H_s=const.$ is adopted.
}
\label{allSFR-fig-const}
\end{figure}

\begin{figure}
\epsscale{1.}
\plotone{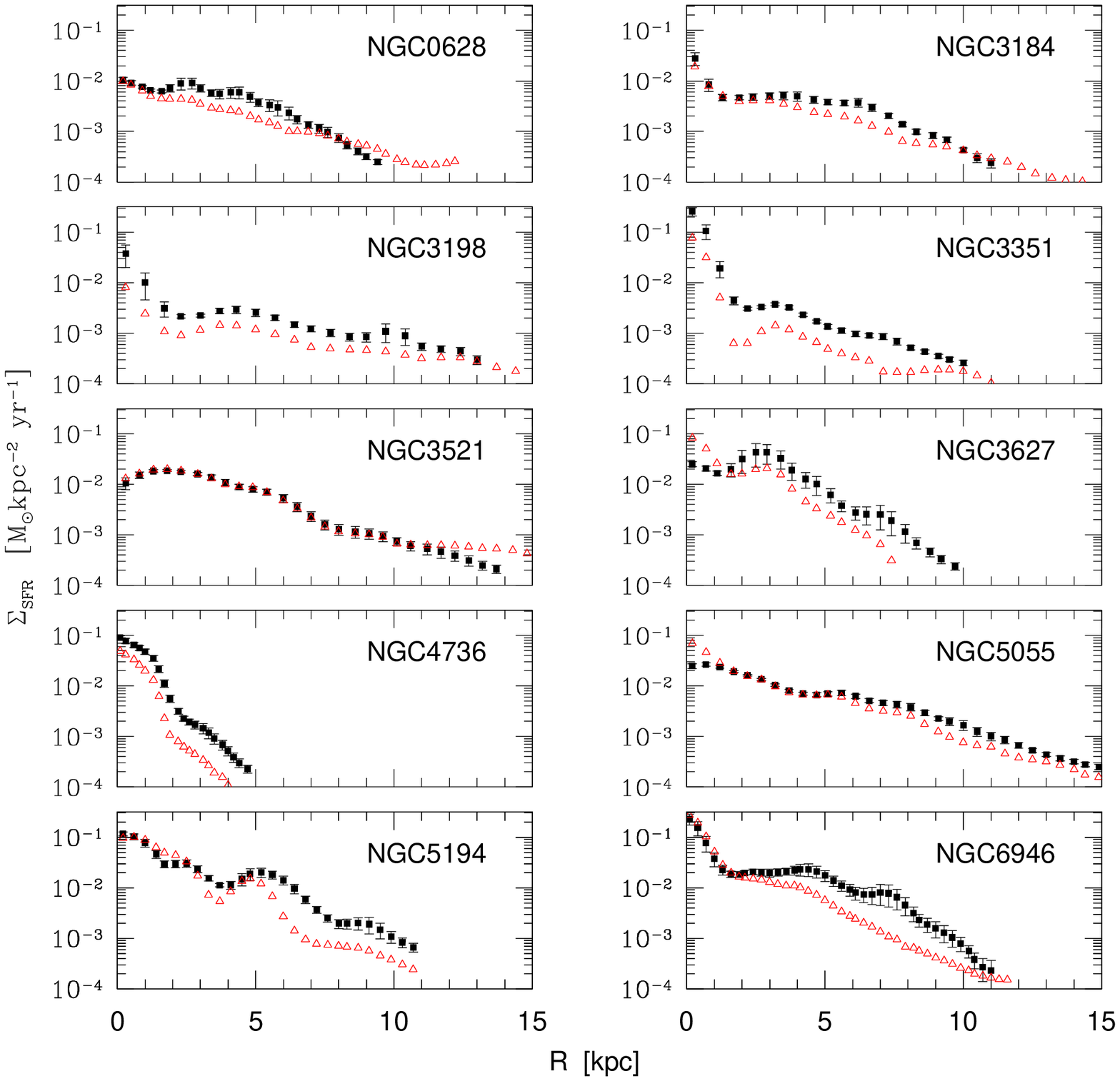}
\caption{Same as in Fig. \ref{allSFR-fig-const}, except
$H_s \propto 1/\Sigma_s$ is adopted for the model.
}
\label{allSFR-fig-flare}
\end{figure}

For the remaining set of ten spiral galaxies in the sample, Figures
\ref{allSFR-fig-const} and \ref{allSFR-fig-flare} show the comparison
between observed and predicted values of $\sigsfr$, for $H_s=const.$
and $H_s \propto 1/\Sigma_s$.  As for NGC 7331, we adopt
the fiducial parameter values $\fwt=0.5$ and $\alpha=5$; no corrections
for departures from Solar gas-to-dust ratio have been made, however.
Overall, the predictions follow the observed profiles fairly well. For
NGC 5055, which, like NGC 7331, is a flocculent galaxy,
quite good agreement is evident ($\simlt 30\%$ differences for 
$H_s=const.$ or $\simlt 50\%$ differences for $H_s \propto 1/\Sigma_s$).  
Again, this is true both for the
inner molecule-dominated region, and for the atomic-dominated region
which lies outside of $\approx 8\kpc$.  When considering
azimuthally-averaged data, it is not surprising that the prediction
should be most accurate for flocculent galaxies: since the dependence
of $\sigsfr$ on $\Sigma$ is nonlinear at high $\Sigma$, the prediction
for the average $\sigsfr$ using the azimuthally-averaged $\Sigma$ can
significantly underestimate 
the average using local values of $\Sigma$ if the gas and
star formation are both strongly concentrated in spiral arms
(see eq. \ref{az-comp} in the Appendix).
Note that both NGC 7331 and NGC 5055 are also observed to 
have particularly clean linear 
relationships between $\sigsfr$ and $\Sigma_{\rm mol}$.

For a few cases (e.g. NGC 3198, NGC 4736, NGC 5194), the
shape of the predicted $\sigsfr$ follows that of the observations in
the atomic-dominated part, but with an offset downward in the overall
magnitude.  From equation (\ref{asymp-SFR}), $\sigsfr \propto
(\fwt/\alpha)^{1/2}$ in the diffuse-gas dominated regime, so that
adopting a number for $\fwt/\alpha$ larger than the fiducial
value we have assumed would shift the predicted $\sigsfr$ upward. Of course, 
it is also possible that observational systematics contribute to this 
offset.  For NGC 4736, the inner molecular-dominated regime has an
offset between predicted and observed $\log(\sigsfr)$ that is similar to the 
outer-disk offset.  Since the predicted 
$\sigsfr$ in inner disks is $\propto \tsf^{-1}$ but independent of 
$\fwt$ and $\alpha$, the similarity of the inner- and outer-disk 
offsets could either mean that
$\tsf^{-1}$ and $(\fwt/\alpha)^{1/2}$ happen to vary together, or that 
the observed star formation rate is systematically overestimated everywhere
(note that this galaxy has other peculiarities in both its star formation
and gas dynamics; see \citealt{2000ApJ...540..771W}). The galaxies NGC 3198
and 3351 have observed values of $\Sigma/\sigsfr$ in their 
inner regions less than the fiducial value $\tsf=2\times 10^9\yr$; 
potentially, the gas surface density may be underestimated if the 
CO-to-H$_2$ conversion factor is too small, or star formation rates may be
overestimated.

For some other 
galaxies (e.g. NGC 0628, NGC 6946, and to a lesser extent, NGC 3184) 
the shape of the predicted and observed $\sigsfr$ differ somewhat.  
The sense of the discrepancy is that the observed $\sigsfr$ is higher than 
the predicted value at intermediate radii.  If some star-forming 
gas is present that 
is not observable either in 21 cm or CO lines, this could 
in part account for the discrepancy. These galaxies also have 
an irregular -- and sublinear on average -- 
relationship between $\sigsfr$ and $\Sigma_{\rm mol}$ as inferred from CO.
This 
might indicate that CO is not a linear tracer of the 
gas in GBCs, that $\tsfgbc$ is not constant in these regions, or 
that age effects in the stellar population are impacting the estimate 
of $\sigsfr$.

\begin{figure}
\epsscale{1.}
\plotone{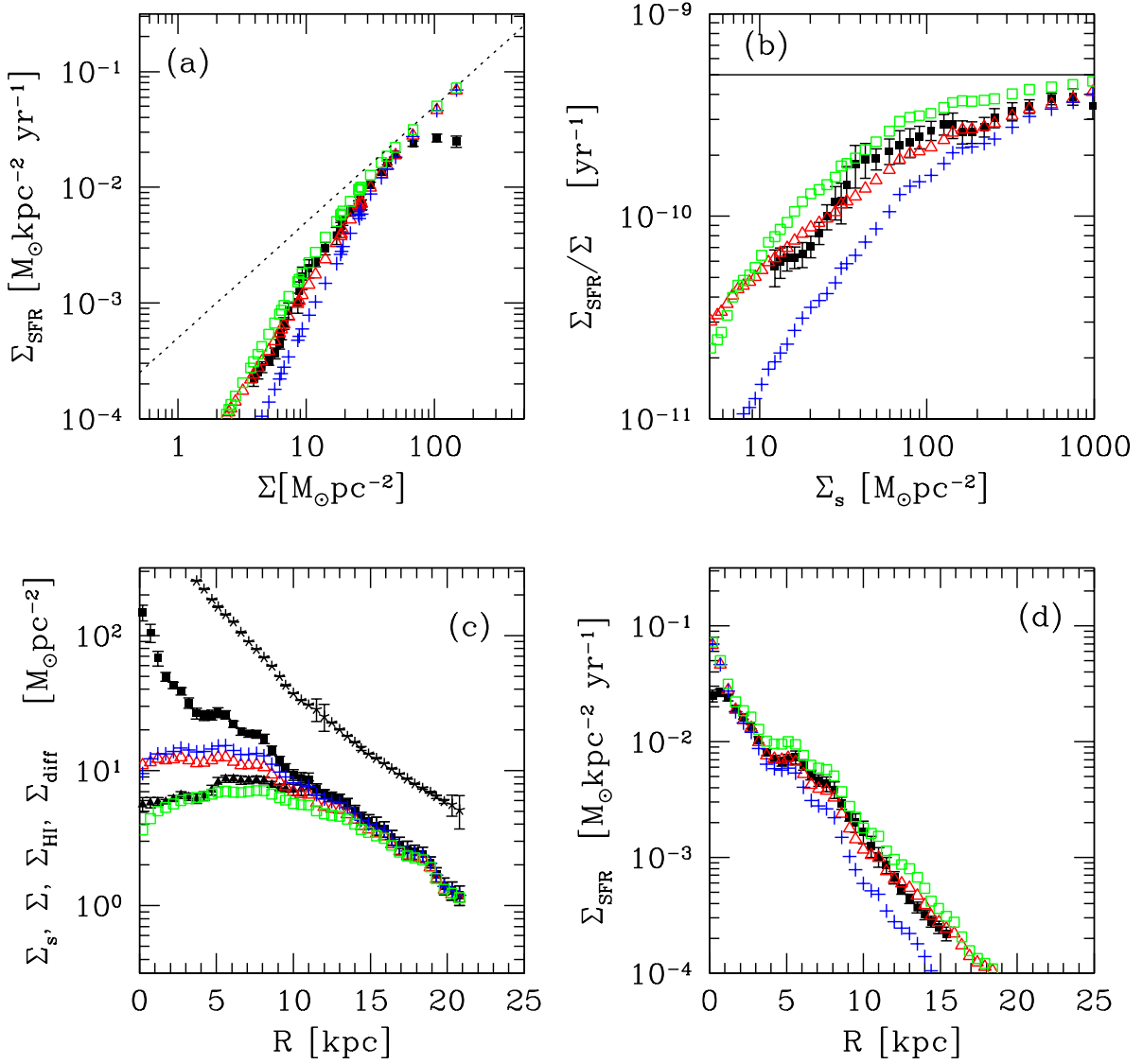}
\caption{Same as in Fig. \ref{NGC7331fig-const}, for the galaxy NGC 5055.
Also included is the comparison with the empirical 
formulae of \citet{BR06} ({\it blue plusses}) and \citet{Ler08} 
({\it green boxes}),
as described in the text.
}
\label{NGC5055fig-const}
\end{figure}

\citet{BR06} obtained an empirical fit relating the molecular-to-atomic
gas mass fractions to a midplane pressure estimate, 
$\Sigma_{\rm mol}/\Sigma_{\rm atom}=(P_{ h}/P_{ h, 0})^{\gamma}$, for 
\begin{equation}
P_h=P_{\rm BR}\equiv \Sigma(2 G \rho_s)^{1/2}  v_g.
\label{BR-press}
\end{equation}
This pressure estimate assumes 
that the stellar disk dominates the vertical gravity, and combines the 
atomic and molecular gas into a single component.  
When combined with 
$\sigsfr=\Sigma_{\rm mol}/\tsf$, this yields 
\begin{equation}
\sigsfr=\frac{\Sigma}{\tsf}\frac{(P_{ h}/P_{ h, 0})^{\gamma}}
{1+ (P_{ h}/P_{ h, 0})^{\gamma}  }.
\label{BR-fit}
\end{equation}
\citet{BR06} adopted a vertical velocity dispersion $v_g=8\kms$, and 
obtained values 
$\gamma=0.92$ and 
$P_{h,0}/k\approx 4\times 10^4 \cm^{-3} \K$
for the fitting constants.

\citet{Ler08} found that a similar relationship fit their sample of spirals,
with $\gamma=0.8$, $P_{h,0}/k=1.7\times 10^4\cm^{-3} \K$, and a pressure
estimate that includes gas self-gravity:
\begin{equation}
P_h=P_{\rm L}\equiv \frac{\pi G \Sigma^2}{2} + \Sigma (2 G \rho_s)^{1/2} v_g;
\label{L-press}
\end{equation}
the adopted value for the vertical velocity dispersion of the composite 
ISM is $v_g=11\kms$.
Note that the gravity of dark matter is not included in either the 
\citet{BR06} or the \citet{Ler08} pressure estimate; dark matter may 
be increasingly important in far-outer disks or in low-surface-brightness 
disks.

For $P_h/P_{h,0}\gg 1$, equation (\ref{BR-fit}) yields the same star 
formation rate as that in the GBC-dominated regime for the present model,
$\sigsfr \rightarrow \Sigma/\tsf$.  For $P_h/P_{h,0}\ll 1$, the limit in 
which atomic gas dominates molecular gas, equation (\ref{BR-fit}) 
yields $\sigsfr \rightarrow \Sigma (P_h/P_{h,0})^\gamma/\tsf$.  
Since $\gamma$ is close to unity for both of these empirical relations,
and $\Sigma$ in the diffuse-dominated limit for both samples is typically 
$\sim 5-10 \Msun \pc^{-2}$ so that $P_h$ varies approximately as $\rho_s^{1/2}$, 
these relationships are similar to the form of our result given in 
equation (\ref{asymp-SFR}) (see also eq. \ref{low-sig-lim}). (Note that a
value of 
$\gamma$ less than 1 partially compensates for varying $\Sigma$ in $P_h$.)
As an example, Figure (\ref{NGC5055fig-const}) presents the comparison 
between our model results and the empirical fits given above, for the 
galaxy NGC 5055.  For this and other galaxies, we find a close correspondence
particularly with the empirical formula of \citet{Ler08} 
(as discussed above, slightly larger $\fwt/\alpha$ than our fiducial choice 
shifts our predicted $\sigsfr$ upward).  The empirical 
formula of \citet{BR06} produces somewhat more rapid decline 
in $\sigsfr$ at low $\Sigma$ (for large radii) than the prediction 
of our model.

\vfil

\section{Summary and Discussion
\label{sec-disc}
}
\bigskip
\leftline
{\it 1. Summary of the physical model --}
In this paper, we have developed a theory for self-regulated star
formation in multiphase 
galactic ISM disks in which stellar heating mediates the feedback.
The fundamental principles we adopt are that for a time-and-space
averaged steady state on $\sim \kpc$ scales, (1) force balance must be
satisfied in the vertical direction, dynamically setting the midplane 
thermal pressure $\Pth$ of the diffuse gas based on the weight 
of overlying material 
(see eq. \ref{pth-eq}); (2) thermal equilibrium must
be satisfied, with the heating rate set by the local
star formation rate (see eqs. \ref{Wol-eq}, \ref{G0-eq}), and with
the two-phase thermal equilibrium pressure $P_{\rm two-phase}$ 
in the diffuse gas equal to the
dynamically-imposed equilibrium pressure $\Pth$; (3)
the star formation rate is controlled by the amount of gas in
GBCs (see eq. \ref{sfr-eq} ),
with the (complementary)
non-self-gravitating amount regulated by the thermal pressure (see
eq. \ref{sigdiff-eq}).  

The set of algebraic
equations embodying the principles above can easily be solved numerically
to obtain $\sigsfr$
as a function of the total gas surface density $\Sigma$ and the midplane
stellar-plus-dark-matter density $\rhoext$, as described in the Appendix. 
An approximate closed-form
solution, representing the key result of this work, is given by equations 
(\ref{asymp-SFR}) and (\ref{SFR-approx}).  In the diffuse-gas
dominated regime, $\sigsfr \propto \Sigma \sqrt{\rhoext}$ (see eq.
\ref{asymp-SFR}).  As a consequence, no single Schmidt-type relation
$\sigsfr \propto \Sigma^{1+p}$ is expected to apply in outer disks,
since $\sqrt{\rhoext}$ need not vary as $\Sigma^p$.

Physically, outer and inner disks are distinguished by which gas component
dominates the mass -- diffuse or self-gravitating.  Where diffuse gas
dominates, the mean pressure and density and hence the cooling rate are 
fixed by the weight of the ISM. 
The amount of self-gravitating, star-forming gas created is
then tuned to provide the needed FUV heating to balance cooling 
at this density: $\sigsfr
\propto P_{\rm th} \propto \Sigma \sqrt{\rhoext}$.  Where self-gravitating 
gas dominates so that $\sigsfr \propto \Sigma$, the specific heating 
rate is fixed. The cooling rate of the diffuse gas depends on 
its density, which is proportional to the surface density $\sigdiff$ 
(and to the vertical gravity);
$\sigdiff$ must therefore adjust until the cooling rate matches the heating 
rate.  The limited surface density observed for \ion{H}{1} gas in the 
central regions of galaxies likely owes at least in part to the constraint
imposed by matching heating with cooling in the diffuse ISM.

\bigskip
\leftline
{\it 2. Connection to previous work --}
Our theory makes use of some of the same concepts -- such as thermal
and dynamic equilibrium -- discussed in previous work, 
but with a different emphasis. 
\citet{1988A&A...205...71P} suggested that star formation is self-regulated
in such a way that the UV radiation it produces 
maintains $P_{\rm max, warm}$ near the thermal pressure
of the gas, and \citet{1991A&A...250...70P} applied this model to the Milky Way 
 by adopting a radial profile for the warm
gas density.
Considering outer galaxies, \citet{1994ApJ...435L.121E}
pointed out that star formation could be strongly suppressed in outer
disks if the midplane pressure falls sufficiently far below $P_{\rm min,
  cold}$ that even locally-compressed regions cannot cool.
\citet{2004ApJ...609..667S}, treating the UV intensity as a fixed
parameter, suggested that star formation would have a threshold
imposed by thermodynamics.  Here, we propose that (low-level)
star formation is able to extend out to large radii in galactic disks because
$P_{\rm min, cold}$ adjusts to follow the outward decline in the midplane
pressure, via a decrease in the UV flux (tracking the decline in the
star formation rate).

In our model, star formation is regulated such
that UV radiation created by young stars heats the disk 
just as much as is needed for the
thermal pressure in the diffuse gas to meet the requirements imposed
by vertical force balance.  The regulation process depends on mass
exchange between self-gravitating and diffuse components of the ISM 
such that star formation at the required rate can take place in 
the bound clouds.  Since the star
formation rate in normal galaxies is proportional to the mass
in gravitationally bound GMCs, we believe
that this self-regulation mechanism is the physical basis for
the relationship between molecular surface density and 
midplane pressure empirically identified by \citet{BR04,BR06}.  
The theoretical relationship we obtain is in fact 
slightly different from the empirical formula of \citet{BR06}
(see also \citealt{Ler08}). 
They found that the ratio of molecular-to-atomic surface
densities is approximately proportional to an estimate of 
the total midplane pressure for a composite ISM 
(see \S \ref{sec-obs} for details), whereas 
here we argue that the surface density of gas contained in GBCs
should be proportional to the midplane thermal pressure in the diffuse gas. 
While physically and mathematically different, the two 
relationships yield quantitatively similar values of 
$\sigsfr$ provided the atomic surface 
density is relatively uniform (as is true in the observations of 
\citealt{BR04,BR06} and \citealt{Ler08}), 
the ratio of thermal to total pressure 
is relatively constant, and the gas in GBCs is mostly molecular.  
In particular, the empirical BR relationship is most sensitive 
in the atomic-dominated limit, where the predicted star formation 
rate  (assuming constant-$\tsf$ in molecular gas) is similar in form
to our theoretical outer-disk law, $\sigsfr \propto \Sigma \sqrt{\rhoext}$.

\bigskip
\leftline
{\it 3. Comparsion to observation, present and future --}
Initial comparisons of predicted star formation rates with observed
values, based on azimuthally-averaged data for disk galaxies, show
quite promising results.  In particular, the predictions follow the
observations very closely throughout the two large flocculent galaxies
NGC 7331 and NGC 5055, out to $1.2 r_{25}$.  In these initial
comparisons, we have not ``tuned'' the model parameters $\alpha$,
$\fwt$, or $\tsf$ at all, but simply adopted the same values for all
the galaxies in the sample.  By adjusting the parameters, closer
agreement between the model prediction and observations can be
obtained in several cases, by shifting the overall normalization of
$\sigsfr$.  Adjusting the prescription for converting stellar surface 
density to volume density can also yield a closer match to the data.  

Current data sets exist that make it possible to extend
the present comparisons in several ways, including using $\sim $kpc
resolution maps (also including local variations of the gas-to-dust
ratio) rather than azimuthally-averaged data, and considering dwarf 
galaxies.  For far-outer disks, the necessary averaging scale
is likely to increase, due to the flaring of the disk.
It will be interesting to test whether galaxies 
with strong spiral structure, when examined locally, are consistent 
with the steady-state theory developed here, or whether transient effects
within spiral arms are too rapid for (quasi-)equilibrium to be attained.

In the analysis of \S 2, we employ several parameters (e.g. $\fwt$,
$\alpha$, $\tsf$, and the ratio of Solar neighborhood pressure 
to star formation rate), 
adopting fiducial values that are
based on current observations and/or theoretical work.
As more
detailed ISM information becomes available from extragalactic
observations (such as local values of the gaseous vertical velocity
dispersion, and the proportions of atomic gas in warm and cold
phases), it will be possible to assign observed values rather than 
adopted parameters as inputs for predicting $\sigsfr$ 
within individual galaxies. 
With local measurements of the stellar
vertical velocity dispersion in the outer parts of 
individual face-on galaxies, 
it will be possible to obtain more
direct estimates of the stellar midplane density, rather than simply adopting 
a prescription for the stellar scale height to obtain $\rho_s$ from 
$\Sigma_s$.  From surveys of edge-on galaxies, it will also be possible
to obtain accurate measurements of the correlations of stellar disk flaring 
with other properties, that could then be applied to more face-on systems
statistically.
With more detailed data sets,
it will be possible to test consituent elements of the theory
-- including equations (\ref{mom-eq}) and (\ref{SFR-P-eq}) -- as well
as the overall prediction for $\sigsfr$.  

\bigskip
\leftline
{\it 4. Opportunities for numerical modeling --}
It is of considerable interest to test 
the idealizations of this theory, as well as its predictions, via detailed
numerical simulations. Numerical models must have sufficiently fine
spatial grids ($\simlt\pc$) that the vertical direction is well
resolved, must include heating and cooling such that warm and cold
phases are present, and must model energetic feedback (leading to both
heating and turbulent driving) from star formation in self-gravitating clouds.  
Because turbulent mixing at extremely small scales ($\ll \pc$)
can affect $\fwt$, it is important to assess this effect using very high 
resolution simulations (e.g. \citealt{2005A&A...433....1A}).

With simulations, the turbulent velocity dispersion and warm/cold
fractions in the diffuse medium can be self-consistently calculated
(cf. \citealt{KO09a}), such that the dependence of the ISM ``state''
parameters $\alpha$ and $\fwt$ on the (input) galactic environment
``variables'' ($\Sigma$, $\rho_s$, $\rhodm$, $Z$) 
and (derived) star formation rate $\sigsfr$ can be assessed.  
The star formation timescale $\tsf$ in GBCs 
and the ratio of mean thermal pressure
(or mean FUV intensity) 
to $\sigsfr$ are also in principle calculable theoretically, 
although the dependence on the fundamental environment 
variables may be fairly complex.

While we have not discussed here exactly how the formation of GBCs
takes place, this
can in principle be strongly affected by the angular momentum of the
disk, with disks having very low or high Toomre $Q$ departing from
observed star formation and molecular/atomic (or GBC/diffuse) 
relations at a given
$\Sigma$ and $\rho_s$ \citep{KO09a,KO09b}.  Normal galaxies have had
sufficient time to evolve (lowering $\Sigma$ and raising $\rho_s$ by
converting gas into stars) that the fundamental dependence of
$\sigsfr$ on angular momentum and shear may, however, 
not be evident in practice.  Thus,
exploring a wide range of types of observed systems, and testing both
realistic and unrealistic galaxy models with numerical simulations,
will be important for revealing the processes that control star
formation at the most fundamental level.
 
\bigskip
\leftline
{\it 5. Limitations and prospects --}
By adopting a fixed value of the star formation timescale $\tsf$ in
self-gravitating clouds, the present model is limited to the regime in
which star-forming clouds have ``normal'' properties, similar to those
observed in the disks of Local Group galaxies
(e.g. \citealt{She08,Bol08}).  In particular, it is not applicable to
galactic center regions or starbursts where $\Sigma$ exceeds the
typical surface density $\sim 100\Msun\pc^2$ of individual mid-disk
GMCs.  In such high-$\Sigma$ 
regions, molecular gas completely dominates the ISM,
but because the average density 
is higher than in mid-disk GMCs -- so that the 
gravitational time
$\propto \rho^{-1/2}$ is shorter, the star formation rate per unit 
mass is expected to be higher than it is for ``normal'' GMCs (consistent with
observations).  In detail, the star formation rate per unit mass in 
bound clouds 
in high-$\Sigma$ regions is also expected to depend on the turbulence
level (which is higher in starbursts) and whether star-forming clouds
are collapsing or marginally bound
(cf. \citealt{KM05,2009arXiv0911.1795R,2009arXiv0907.0248P}).  
Thus, from a combination of these effects, $\sigsfr/\siggbc$ is not 
expected to be constant in high-$\Sigma$ regions, and the present theory 
should not be applied there.

We note that low-metallicity systems, where GBCs
are less self-shielded and have larger atomic-to-molecular
ratios than ``normal'' GMCs of the same size and mass, may also have
values of $\sigsfr/\siggbc$ different from the constant value
$\tsf^{-1}=(2\times 10^9)^{-1}\yr$ we have adopted (based on the CO
observations of \citealt{Bigiel08}).  In bound clouds with a large
atomic-gas proportion (and with the carbon mostly atomic rather 
than in CO), the mean temperature will be larger than that
in primarily-molecular clouds.  Because of the lower internal Mach
number, turbulent compression would be less extreme than in colder,
primarily-molecular, clouds, which could affect the fraction of gas
that is able to collapse and make stars.  Additional observational and
theoretical work is needed to evaluate how the star-forming
efficiencies of bound clouds depend on relative amounts of cold-atomic
vs. molecular gas, and also to explore whether the masses and/or total
column densities of star-forming bound clouds in low-metallicity
regions (far-outer disks of spirals, and dwarf galaxies) 
differ systematically from the properties of ``normal'' GMCs.
Observationally, a difficulty (particularly in low metallicity 
regions) is that significant gas can be ``dark'' \citep{2005Sci...307.1292G} 
-- i.e. not observable either in HI (because the hydrogen is molecular) or 
in CO (because the carbon is atomic).  For Solar metallicity and clouds 
with $A_V\sim 8$, the fraction of ``dark'' gas is expected to be only 
$\sim 0.3$ \citep{2010ApJ...716.1191W}.

The present model does not address radiative transfer effects
explicitly.  In particular, we have assumed that the optical depth
through the diffuse gas is modest, such that the mean UV intensity and
therefore the thermal pressure is approximately 
proportional to the local $\sigsfr$
(see eqs. \ref{Wol-eq} - \ref{SFR-P-eq}).  Although this approximation
becomes invalid where $\sigdiff \simgt 20/Z'\Msun\pc^{-2}$ (for $Z'$ the 
metallicity relative to Solar), 
typically $\Sigma \gg \sigdiff$ by this point (see
e.g. Fig. \ref{fig-examp}), so that changes in the FUV intensity and hence 
$P_{\rm two-phase}$ and $\Pth$
would not significantly affect the predicted $\sigsfr$.  An accurate
determination of $\sigdiff$ where $\Sigma Z'$ is high would, however,
require an explicit radiative transfer calculation to assess $G_0'$ for
a given value of $\sigsfr$.  This would be affected both by the amount and
vertical distribution of diffuse gas, and by radiative transfer within
star-forming clouds themselves.  Radiative transfer effects also become 
important in the far outer regions of galaxies, where heating from nonlocal
UV can exceed the local contribution.
Results of radiative transfer 
models could in principle be tested by comparison to 
multiwavelength IR observations, since the dust temperature is sensitive
to $G_0'$ (i.e., to the mean UV intensity $J_{\rm FUV}$) 
\citep{2007ApJ...663..866D}. Here, we have not attempted to
address these issues, but instead we have simply adopted an 
empirical Solar-neighborhood value for the ratio of $J_{\rm FUV}$ to 
$\sigsfr$ to calibrate our relationships.  

Determination of the relative proportions of atomic and molecular
(and, for low $\Sigma$, ionized) gas also depends on radiative
transfer.  For GBCs, the solution
for spherical clouds of \citet{KMT09} and \citet{2010ApJ...709..308M}
predicts the molecular-to-atomic ratio is 
$(M_{\rm atom}/M_{\rm mol})_{\rm cloud}\approx [Z'^{0.8}
  (N_{\rm H,cloud}/1.8\times 10^{21}\cm^{-2})-0.7]^{-1}$; if clouds 
either have fixed 
total hydrogen column density or there is a known relationship between
the mean surface density of this gas component 
$\siggbc$ (averaged over $\sim \kpc$ scales) 
and the column density of individual clouds, then this could
be used to compute the contributions to $\Sigma_{\rm mol}$ and
$\Sigma_{\rm atom}$ from $\siggbc$.  
The diffuse gas could also be partly molecular; using 
the results of \citet{2010ApJ...709..308M} for a slab of cold 
gas illuminated on both sides, a layer
begins to become molecular when 
$\Sigma_{\rm cold} \simgt 11 \Msun\pc^{-2}/[Z'^{0.8}]$.

Finally, we note that simple models of the kind we have developed here
-- if validated  by detailed numerical simulations and confirmed by
observations -- potentially 
provide a valuable tool for studies of
galaxy evolution. We caution, however, that careful appraisal of
metallicity effects will be required before applying this model (or 
a refined version) to
star formation in galaxies at high redshift.  Our results could also
potentially be adapted to provide subgrid ISM/star formation
prescriptions for use in cosmological simulations of galaxy formation,
but this too should be approached with care since the ratio of
diffuse to gravitationally-bound gas is computed not for a local
three-dimensional zone but for a vertically-integrated disk. 
Even with a subgrid model, the scale height of the diffuse warm+cold 
ISM ($H\sim \sigma_z^2/g_z\sim \sigma_z/\sqrt{4\pi G\rho_{\rm tot}}$) 
must be resolved by several zones 
in order to obtain an accurate estimate for the midplane 
pressure, which controls the amount of star-forming gas. Because
formation of gravitationally-bound star-forming clouds depends on
non-local dynamical processes, simulations that are either resolved
sufficiently to capture gravitationally-induced vertical motions, or
are completely vertically-unresolved with a suitable prescription for
balance of ISM components, can represent the relevant physics more faithfully
than simulations which resolve the disk with just a few zones.

%\begin{

\acknowledgements

The work of ECO was supported in part by fellowships from the Miller
Institute at U.C. Berkeley and the John Simon Guggenheim Foundation,
and by grants AST-0908185 from the National Science Foundation and
NNG05GG43G from NASA.  The work of CFM was supported in part by grant
AST-0908553 from the National Science Foundation.  CFM also
acknowledges the support of the Groupement d'Int\'er\^et Scientifique
(GIS) ``Physique des deux infinis (P2I)''.  Support for AKL was
provided by NASA through Hubble Fellowship grant HST-HF-51258.01-A
awarded by the Space Telescope Science Institute, which is operated
by the Association of Universities for Research in Astronomy, Inc.,
for NASA, under contract NAS 5-26555.
We are grateful to Bruce
Elmegreen, Rob Kennicutt, and Antonio Parravano for helpful comments
on the manuscript, and to the referee for a thoughtful and
constructive report.

%\end{acknowledgements}

\appendix

\section{Appendix}

Here, we provide an explicit formula that is solved numerically to
obtain the fractions of diffuse and gravitationally-bound gas. 
We show how this formula leads to
the equations (\ref{asymp-SFR}) and (\ref{SFR-approx}) given in the
text for the star formation rate.  We also discuss how the 
galactic environment variables ($\Sigma$, $\rhoext$, $Z$) and the 
ISM model parameters 
control the transition between the diffuse-dominated and
GBC-dominated regimes, and how presence of both regimes in a given annulus 
(due to spiral structure) affects estimates of the star formation rate.
Finally, we provide a more stringent upper limit on
$\sigdiff$ than the limit given by equation (\ref {Siglim-eq}) in the
text.

We start with the diffuse-gas thermal equilibrium 
equation (\ref{SFR-P-eq}) relating the thermal pressure 
to the star formation rate $\sigsfr=\siggbc/\tsf$, which may be expressed 
as 
\begin{equation}
\Pth = \frac{1}{\phi_d} \frac{  P_{\rm th, 0}}{\Sigma_{SFR,0}}
\frac{\siggbc}{\tsf}. 
\label{Pth-eq}
\end{equation}
Here, $P_{\rm th,0}$ and $\Sigma_{SFR,0}$ are the 
Solar-neighborhood thermal pressure of diffuse gas and star 
formation rate that we adopt for our normalizations, and 
\begin{equation}
\phi_d \equiv \frac{1}{4} 
\left[1+ 3 \left(\frac{Z_d'\Sigma}{\Sigma_0} \right)^{0.4}  \right]
\end{equation}
is defined such that it is equal to unity for $\Sigma=\Sigma_0$, the
Solar neighborhood diffuse-gas surface density. 
The ratio of 
$P_{\rm th,0}$ to the mean local FUV intensity $J_{\rm FUV,0}$ is computed 
theoretically \citep{Wol03}, so that Solar neighborhood 
observations effectively 
enter the model through the ratio $J_{\rm FUV,0}/\Sigma_{\rm SFR,0}$.
Using Solar neighborhood 
values $P_{\rm th,0}/k=3000 \K \pcc$ and $\Sigma_{\rm SFR,0}=2.5\times 
10^{-9} \Msun \pc^{-2} \yr^{-1}$ with $\tsf=2 \times 10^9\yr$ as determined 
from extragalactic studies, this yields
$\Pth/k = 600 \K \pcc \phi_d^{-1}(\siggbc/\Msun \pc^{-2})$ for the 
relation between thermal pressure and 
the surface density of gas in gravitationally-bound (star-forming) 
clouds.\footnote{Note 
that inserting $\siggbc \simlt 2 \Msun \pc^{-2}$, as indicated by 
Solar neighborhood observations of molecular gas
\citep{1987ApJ...322..706D,1988ApJ...324..248B,2001ApJ...547..792D,2006ApJ...641..938L,2006PASJ...58..847N}, would yield a pressure a factor of $\sim 2$
below the observed local value.  This simply reflects the fact that 
$\Sigma_{SFR,0}\simlt 10^{-9} \Msun \pc^{-2} \yr^{-1}$ if 
$\siggbc \simlt 2 \Msun \pc^{-2}$ and $\tsf=2 \times 10^9\yr$, 
whereas observational estimates of 
$\Sigma_{SFR,0}$
are higher by a factor $\sim 2$ 
\citep{2001AJ....121.1013B,2002A&A...390..917V,Fuchs2009}.  
It is uncertain whether 
this discrepancy is the result of an underestimate of the local $\siggbc$,
an overestimate of the local $\sigsfr$, or a lower local value for 
$\tsf$ than the mean extragalactic value.} 

Next, we define 
\begin{eqnarray}
\sigdiff &\equiv& x\Sigma, \\
\siggbc &\equiv& (1-x)\Sigma,
\end{eqnarray}
for the diffuse-gas and GBC surface densities, substituting these expressions
together with equation (\ref{Pth-eq}) for $\Pth$ in equation (\ref{sigdiff-eq})
to obtain:
\begin{equation}
x \Sigma= \frac{\pi G  \Sigma_h \Sigma(1-x)}
{\pi G \Sigma(1-x) + \left[(\pi G \Sigma)^2(1-x)^2 + 
(\pi G)^2 \Sigma \Sigma_h(1-x)
 + 8\pi G\zeta_d c_w^2 \fwt \alpha \rhoext  \right]^{1/2}  }.
\label{x-eq-prelim}
\end{equation}
Here, we have introduced 
\begin{eqnarray}
\Sigma_h&\equiv& \frac{2\alpha  P_{\rm th, 0}}{\pi G \phi_d \Sigma_{SFR,0} \tsf}
\label{Sigma_h-eq}
\\
&=&91 \Msun \pc^{-2} \phi_d^{-1}\left(\frac{\alpha}{5}\right) 
\left(\frac{P_{\rm th,0}/k}{3000 \K \cm^{-3}}\right) 
\left(\frac{\Sigma_{SFR,0}}{2.5\times 10^{-9} \Msun \pc^{-2}\yr^{-1}}\right)^{-1}
\left(\frac{\tsf  }{2\times 10^9 \yr  } \right)^{-1}.\nonumber
\end{eqnarray}
Since $P_{\rm th,0}/(\phi_d \Sigma_{\rm SFR,0} \tsf)=\Pth/\siggbc$ 
and $\alpha \Pth$ is the total (effective) midplane pressure, $\Sigma_h$
is the value that $\Sigma$ would have to attain in order for the total 
pressure to equal $\pi G \Sigma \siggbc/2$; that is, for the pressure
to be dominated by the gravity of the gas.  

Next, we define
% \begin{equation}
% S\equiv \frac{8\zeta_d c_w^2 \alpha \fwt}{\pi G}\frac{ \rhoext}
% {\Sigma^2},
% \end{equation}
\begin{eqnarray}
S&\equiv& \frac{8\zeta_d \alpha \fwt c_w^2 }{\pi G}\frac{ \rhoext} {\Sigma^2}\\
&=&31 \left(\frac{\alpha}{5}\right)\left(\frac{\fwt}{0.5}\right)
\left(\frac{\rhoext}{0.1 \Msun \pc^{-3}}\right)
\left(\frac{\Sigma}{10 \Msun \pc^{-2}}  \right)^{-2},
\end{eqnarray}
which measures the relative importance of external gravity
(stellar-disk + dark matter) to gas self-gravity in setting the diffuse-gas 
pressure.  We note that if $\rhoext$ is dominated by stars with
$\rho_s=\Sigma_s/(2H_s)=\pi G \Sigma_s^2/(2 v_{z,s}^2)$,
\begin{equation}
S=4 \zeta_d \frac{\langle v_{\rm th}^2 + v_t^2 \rangle  }{v_{z,s}^2 }
\frac{\Sigma_s^2  }{\Sigma^2  },
\end{equation}
which is $\propto (Q_{gas}/Q_{s})^2$ in terms of the stellar and gas Toomre 
parameters (cf. equation 4 in \citealt{KO09b}).  
Using fiducial values $\alpha=5$ and $\fwt=0.5$, 
$S\sim 16$ in the Solar neighborhood.

We can now re-express equation (\ref{x-eq-prelim}) in terms of the 
dimensionless variables $S$ and 
\begin{equation}
w\equiv \frac{\Sigma  }{\Sigma_h},
\end{equation}
yielding
\begin{equation}
\frac{1}{x} = w \left\{ 1 + \left[1 + \frac{1  }{(1-x)w  } +
\frac{S}{(1-x)^2}  \right]^{1/2}\right\}.
\label{x-eq}  
\end{equation}
Given values of $\rhoext$ and $\Sigma$ in a galaxy, the variables $S$ and $w$
are set, and we can solve the (cubic) equation (\ref{x-eq}) for $x$ 
numerically. 
The root $x$ is bounded by 0 and 1, so we use the 
bisection method. Given a solution for the diffuse-gas fraction 
$x$, the star formation rate is then 
\begin{equation}
\sigsfr=(1-x)\frac{\Sigma}{\tsf}.
\label{SFR-eq}
\end{equation}
We use this numerical solution to obtain the star formation rates for both our 
idealized galaxy (Fig. \ref{fig-examp}), and for the comparison of the 
model prediction to the observed star formation rates (Figs. 
\ref{NGC7331fig-const} -  
\ref{NGC5055fig-const}).
%\ref{allSFR-fig-flare}).

One
important limit is that in which the ISM is dominated by diffuse gas, in 
which case $x\rightarrow 1$, and equation (\ref{x-eq}) yields
% \begin{eqnarray}
% 1-x &\rightarrow& w\frac{1+\left[1+ 4 S\right]^{1/2}}{2}\\
% &\approx& w(1 + S^{1/2}).
% \end{eqnarray}
\begin{equation}
 1-x \rightarrow w\frac{1+\left[1+ 4 S\right]^{1/2}}{2}
\approx w(1 + S^{1/2}).
\label{low-sig-lim}
\end{equation}
When multiplied by $\Sigma$ and divided by $\tsf$, this yields equation
(\ref{asymp-SFR}) of the text.

Equation (\ref{low-sig-lim}) requires $w(1+ S^{1/2})\ll 1$ for self-consistency,
whereas $x \rightarrow 0$ in equation (\ref{x-eq}) 
requires $w(1+ S^{1/2})\gg 1$.
An approximate solution for $1-x$ allowing for both limits is 
\begin{equation}
\frac{1}{1-x} \approx \frac{1}{w(1+ S^{1/2})} + 1.
\label{gbc-frac}
\end{equation}
The inverse of this, when multiplied by $\Sigma$ and divided by $\tsf$, 
yields the approximation for the star formation rate given by 
equation (\ref{SFR-approx}) of the text.  
Since $w \simlt 1$ for the regions
we are considering in this paper, it is generally the value of $w S^{1/2}$
(which depends on $\rhoext$ but not on $\Sigma$) that determines which 
star formation regime holds.  
The approximation for $1-x$ given in equation (\ref{gbc-frac}) is good to 
within 11\%  for $S\ge 10$ and $w\ge 0.01$; for $w\le 0.2$ and 
$0.1\le S \le 10$,
a better approximation (good to within 12\%) is obtained by 
using $0.5[1+ (1+ 4S)^{1/2}]$ instead of $1+ S^{1/2}$ (cf. eq. 
\ref{low-sig-lim}).

Using equation (\ref{gbc-frac}) and 
the definitions of $w$ and $S$, 
the overall timescale for gas to be converted to stars is given by 
\begin{eqnarray}
t_{\rm con}&\equiv& \frac{\Sigma}{\sigsfr} = \frac{\tsf}{1-x} \\
%\nonumber \\
&\approx& \frac{P_{\rm th,0}}{\phi_d \Sigma_{\rm SFR,0}}
\left(\frac{\alpha  }{2\pi G \zeta_d c_w^2 \fwt \rhoext}\right)^{1/2}
+\tsf,
\label{tcon-eq}
\end{eqnarray}
where $S\gg 1$ is assumed for the latter expression.
This timescale will be set by whichever of the two terms is larger. The
first term is proportional to the vertical oscillation time, 
$\sqrt{\pi/(G\rhoext)}$, which controls how fast cold cloudlets can sink to 
the midplane and form GBCs (for our fiducial parameter choices, 
this term in $t_{\rm con}$ is 39 times the vertical 
oscillation time).  The second term is the characteristic time
for gas within GBCs to form stars.

Quantitatively, 
the transition between the diffuse-dominated and GBC-dominated cases occurs
where $x=1/2$. From equation (\ref{x-eq}), this yields the condition
\begin{eqnarray}
w_{1/2} &=& \frac{2}{3}(1- S w^2)=  
\frac{2}{3}\left(1- \frac{8 \zeta_d\alpha \fwt c_w^2 \rhoext}{\pi G \Sigma_h^2}
\right), 
\end{eqnarray}
where $\Sigma_h$ is given in equation (\ref{Sigma_h-eq}).
Since the right-hand side depends only $\rhoext$, this gives a value 
for the transition surface density
$\Sigma_{1/2}=w_{1/2} \Sigma_h$ as a function of $\rhoext$.  
Taking the fiducial
parameter choices, $S w^2 =0.37 (\rhoext/0.1 \Msun\pc^{-3})$.
For a given value
of $\rhoext$, the ISM will be diffuse-dominated if $\Sigma<\Sigma_{1/2}$, and 
GBC-dominated if $\Sigma>\Sigma_{1/2}$.  For example, at the Solar circle
where $\rhoext=0.05 \Msun\pc^{-3}$, the transition from diffuse-dominated to
GBC-dominated would occur 
at $\Sigma_{1/2}=0.54 \Sigma_h \sim 50 \Msun \pc^{-2}$.
We note that if $\rhoext$ is large enough that $S w^2>1$, 
then $x<1/2$ (i.e. $\siggbc>\sigdiff$) regardless of the value of $\Sigma$.

At a given galactocentric radius, if the surface density satisfies either 
$\Sigma\ll \Sigma_{1/2}$  or 
$\Sigma\gg \Sigma_{1/2}$ at all azimuthal angles, then
a single star formation regime applies, and the azimuthally-averaged star 
formation rate can be obtained from the azimuthally-averaged gas 
surface density.  If, however, there is a transition from 
$\Sigma<\Sigma_{1/2}$ in interarm regions to $\Sigma>\Sigma_{1/2}$ 
in spiral arm regions, then the
star formation regime changes from diffuse-dominated to GBC-dominated, and
the prediction of $\sigsfr$ based on $\langle \Sigma \rangle$
would depart from the true value due to nonlinearities.  
Equation (\ref{gbc-frac}), when multiplied by $\tsf$ and 
evaluated using the value of $w$ in the arm  
gives $t_{\rm con,arm}\equiv\Sigma_{\rm arm}/\Sigma_{\rm SFR,arm}$ in the arm gas:
\begin{equation}
t_{\rm con,arm} \approx 
\frac{\tsf}{w_{\rm arm}+ w_{\rm arm}S_{\rm arm}^{1/2}} + \tsf
\end{equation}
(note that $S\gg 1$ does {\it not} hold in arms, so that eq. 
\ref{tcon-eq} should not be used; $w_{\rm arm}$ may however be 
$\simgt 1$).
An analogous expression holds for 
$t_{\rm con,ia}\equiv\Sigma_{\rm ia}/\Sigma_{\rm SFR,ia}$ in 
the interarm region using $w_{\rm arm} \rightarrow w_{\rm ia}$ and
$S_{\rm arm} \rightarrow S_{\rm ia}$. 
Letting $f_{\rm arm}$ be the mass fraction in the arm in a given annulus, 
the star formation rate in the annulus using the arm and interarm conditions
separately is
\begin{equation}
\Sigma_{\rm SFR,arm+ia}=\frac{  \langle \Sigma \rangle }{t_{\rm con,ia}  }
\left[1 +
f_{\rm arm}\left(\frac{t_{\rm con,ia}}{t_{\rm con,arm}} - 1  \right)
\right].
\label{SFR-arm-ia}
\end{equation}

At a given radius, $w S^{1/2} \propto \rhoext^{1/2}$ varies by $\simlt 10\%$ from 
spiral perturbations,  so that  
$w_{\rm ia}S_{\rm ia}^{1/2} \approx w_{\rm arm}S_{\rm arm}^{1/2}
\rightarrow w S^{1/2}$.  
For most regions of interest, 
$w_{\rm ia}\ll w S^{1/2} \ll 1$ so that 
$t_{\rm con,ia}\approx \tsf (w S^{1/2})^{-1}$.
Equation (\ref{SFR-arm-ia}) can be compared to the 
star formation rate that would be estimated using 
$\langle \Sigma \rangle$ in equation (\ref{gbc-frac}), which yields 
$\Sigma_{\rm SFR, az}= \langle \Sigma \rangle (1-x)_{\rm az}/\tsf 
\approx  \langle \Sigma \rangle w S^{1/2}/\tsf 
\approx \langle \Sigma \rangle/t_{\rm con,ia}$ 
(assuming $\langle w\rangle \ll w S^{1/2}$). 
Taking the ratio of $\Sigma_{\rm SFR, arm+ia}$ to $\Sigma_{\rm SFR,az}$, we obtain
\begin{eqnarray}
\frac{\Sigma_{\rm SFR, arm+ia }}{\Sigma_{\rm SFR,az}}&\approx& 
1 + f_{\rm arm}
\left(
\frac{w_{\rm arm}  }
{wS^{1/2}[w_{\rm arm} + wS^{1/2} +1]  }
\right).
\label{az-comp}
\end{eqnarray}
Since the term in 
parentheses is typically order-unity, the true star formation rate 
(i.e. $\Sigma_{\rm SFR, arm+ia }$) can 
be considerably larger than the estimate $\Sigma_{\rm SFR,az}$ 
based on the annular azimuthal average $\langle \Sigma \rangle$
if the gas is highly concentrated in the arms. 
 
Finally, we consider the upper limit on the diffuse-gas surface density.
Taking the inverse of equation (\ref{x-eq}) and multiplying by $\Sigma$, we 
have for the diffuse-gas surface density
\begin{eqnarray}
\sigdiff &=&  
\frac{\Sigma_h}{1 + \left(1 + \frac{\Sigma_h }{\Sigma(1-x)} + 
\frac{S}{(1-x)^2}\right)^{1/2}}\\
&<&\frac{\Sigma_h}{1 + \left(1 + \frac{\Sigma_h }{\Sigma} + S\right)^{1/2}}.
\label{sigdiff-lim}
\end{eqnarray}
The right-hand side of the inequality (\ref{sigdiff-lim}) 
is the limiting value of $\sigdiff$ 
for $x\ll 1$, i.e. the case in which the ISM is dominated by GBCs.
An absolute upper limit $\sigdiff<\Sigma_h/2$ is obtained by taking 
$\Sigma_h/\Sigma, S\rightarrow 0$;  the result 
is given by equation (\ref {Siglim-eq}) of the text.  In practice, 
the terms $\Sigma_h/\Sigma$ and $S$ in the denominator 
of equation (\ref{sigdiff-lim}) are appreciable, so that 
$\sigdiff$ is below $\Sigma_h/2$ by a factor of a few.
% Taken as an estimate for $x$, equation (\ref{gbc-frac}) is good within 
% 5\% for $w\ge 0.01$ and $S\ge 1$.

%\bibliography{ref}

\end{document}